\documentclass{article}

\usepackage{arxiv}

\usepackage[utf8]{inputenc}
\usepackage[T1]{fontenc}

\usepackage{amsmath}
\usepackage{amsfonts}

\usepackage{array}
\usepackage{booktabs}

\usepackage{graphicx}
\usepackage{xcolor}

\usepackage{nicefrac}
\usepackage{microtype}

\usepackage{natbib}
\usepackage{doi}
\usepackage{url}
\usepackage{hyperref}
\setcitestyle{numbers,square,comma}

\usepackage{xcolor}
\definecolor{rarepink}{HTML}{EF3B67}

\title{Rare Disease Classification in Video Capsule Endoscopy}

\author{
\href{https://orcid.org/0009-0006-9871-1266}
{\includegraphics[scale=0.06]{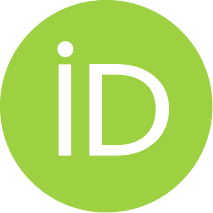}\hspace{1mm}Amit Kumar Patel} \\
Department of Computer Science\\
Biomedical Perception \& Intelligence Lab\\
University of South Dakota\\
Vermillion, South Dakota, USA\\
\texttt{amit.patel@coyotes.usd.edu}
\And
\href{https://orcid.org/0000-0002-8078-6730}
{\includegraphics[scale=0.06]{orcid.pdf}\hspace{1mm}Debesh Jha} \\
Department of Computer Science\\
Biomedical Perception \& Intelligence Lab\\
University of South Dakota\\
Vermillion, South Dakota, USA\\
\texttt{debesh.jha@usd.edu}
}

\date{}

\renewcommand{\shorttitle}{Rare Disease Classification in VCE}

\hypersetup{
pdftitle={Rare Disease Classification in Video Capsule Endoscopy},
pdfsubject={Computer Vision, Medical Imaging, Video Capsule Endoscopy},
pdfauthor={Amit Kumar Patel, Debesh Jha},
pdfkeywords={video capsule endoscopy, rare disease classification, training-free decoding, class imbalance, macro-F1},
}

\begin{document}
\maketitle

\begin{abstract}
Video capsule endoscopy offers a non-invasive way to examine the small bowel, however, the rare but clinically important findings that appear only in less than 1\% of labelled frames, provide very less signal for detection. As a result, while models can achieve high overall performance, they almost completely miss these rare findings. We condition each frame's decision on evidence from the rest of that patient's video, not on the frame's own posterior alone, supplying information a single frozen frame can never contain. Building upon Expected Gain Arbitration, a per-frame, macro-F1-optimal correction over the confusion matrix, we further compose it with video-disease-presence conditional resulting in our decoding method Hierarchical Conditional-Expected-gain (HCE). We evaluate HCE on two independent VCE datasets, Kvasir-Capsule and GALAR, against a static temporal-smoothing baseline and an optimal-transport (OT) frame-level correction. HCE significantly improves rare-class recall over a temporal-smoothing baseline on GALAR (2.7--3.1x; 95\% CI excludes zero in both scopes) and shows a similar, though not statistically confirmed, gain on Kvasir-Capsule (up to 3.4x). Against OT, HCE still improves rare-class recall in all four evaluation settings (+0.016 to +0.029), and additionally improves macro-F1 and MCC.
\end{abstract}

\section{Introduction}
\label{sec:intro}

Video Capsule Endoscopy (VCE) has become a standard tool for non-invasive screening of the small bowel, which is largely inaccessible to conventional endoscopy~\cite{iddan2000wireless}. This process produces videos where each video consists of a large number of highly redundant frames~\cite{smedsrud2021kvasircapsule,lefloch2025galar}. A clinician spends hours going through all the frames to identify the pathological ones, and lesions are frequently missed even by experienced readers~\cite{zheng2012detection}. These pathological frames constitute less than 1\% of the total number of frames in a video~\cite{smedsrud2021kvasircapsule,lefloch2025galar}. Moreover, some pathologies are rare not only within a patient's video, but across the patient distribution. Also, some pathologies carry very little signal, either due to their small region of interest or their high resemblance to dominant classes. Classifiers trained on these datasets face an extreme imbalance in the class distribution, and although they report high overall performance~\cite{srivastava2022focalconvnet,subedi2025cnntransformer}, they are less effective clinically. As a result, flagging these rare pathologies is crucial to make these classifiers clinically useful.

Existing methods to account for class imbalance, such as resampling~\cite{buda2018systematic}, loss reweighting~\cite{cui2019classbalanced,lin2017focal,cao2019ldam}, and decoupled or two-stage training~\cite{kang2020decoupling,cao2019ldam}, require retraining the model, which is impractical or time-consuming once the model is deployed. Other ways of learning robust features~\cite{khosla2020supcon,ranabhat2025shapebias} struggle because the few rare samples get washed away by the dominant classes. Prototype-based classification~\cite{snell2017prototypical} might slightly improve recall, but in our experiments it reduces the discriminative power of the model. Self-supervised pretraining~\cite{oquab2023dinov2} and fine-tuning improve overall performance; however, they cannot lift rare-class recall due to the same limited-signal problem. Post-hoc corrections applied at inference, like logit adjustment~\cite{menon2021logitadjustment}, cost-sensitive thresholding~\cite{koyejo2014consistent,narasimhan2015consistent,parambath2014}, and optimal transport~\cite{jin2023optransport,cuturi2013sinkhorn}, are training-free; however, they cannot escape the imbalance beyond a point due to limited per-frame information. Information beyond a frame's own content is therefore essential to provide richer context~\cite{wang2016tsn,son2022smallbowel}.


We propose Hierarchical Conditional-Expected-gain (HCE), a decoding method aimed at maximizing rare-class recall, which composes a video-level disease-presence conditioning term with Expected Gain Arbitration (EGA), a per-frame correction that reassigns a frame only when doing so increases expected macro-F1, grounded in plug-in consistency results for non-decomposable metrics~\cite{narasimhan2015,koyejo2014consistent}. We validate both our methods on two VCE datasets, GALAR~\cite{lefloch2025galar} and Kvasir-Capsule~\cite{smedsrud2021kvasircapsule}, and report the mean and standard deviation across 3 seeds (single models) and 2 folds (ensembles). As frames within a video are correlated, standard per-frame significance tests overstate confidence. Hence, we use a per-video cluster bootstrap~\cite{field2007bootstrapping} along with a practical significance floor ($|\Delta| \geq 0.005$, or $0.003$ for rare recall), to avoid reporting trivially small effects that are significant only because of the large frame count. Our contributions are as follows:

\begin{enumerate}
\item We introduce HCE, a decoding rule composing a macro-F1-optimal per-frame correction with an offset conditioned on video-level disease presence. It runs on a frozen backbone with no retraining and improves rare-class recall over both a temporal-smoothing baseline and an optimal-transport correction in all four settings, by $1.7$--$3.4\times$ and $1.2$--$1.9\times$ respectively (Kvasir-Capsule $0.064$, $0.101$; GALAR $0.059$, $0.060$).

\item We characterise when video-level conditioning helps: alone it contributes only where per-video presence varies (Kvasir-Capsule $+0.010$, $+0.045$ over temporal smoothing) and is near-inert where videos contain most classes (GALAR $+0.003$, $+0.004$). Composed with EGA, it gives the best rare recall in three of four settings and the highest mean across them, $16\%$ above the per-frame correction alone, $39\%$ above optimal transport, $68\%$ above conditioning alone and $2.6\times$ temporal smoothing.

\item We test significance with a per-video cluster bootstrap that resamples whole videos, since frames within a video are strongly correlated and treating them as independent inflates significance. Of forty method--metric comparisons, five resolve, all on GALAR: rare recall against temporal smoothing in both scopes ($+0.037$, 95\% CI $[0.010, 0.068]$; $+0.041$, $[0.005, 0.067]$), and against optimal transport, macro F1 in both scopes ($+0.022$, $[0.006, 0.036]$; $+0.025$, $[0.003, 0.044]$) and MCC on single models ($+0.094$, $[0.004, 0.168]$). Kvasir-Capsule supplies 25 test videos per fold with a rare class present in roughly seven, so its intervals are wide and none excludes zero.
\end{enumerate}

\section{Related Work}
\label{sec:related_work}
We place our work at an intersection of class-imbalanced learning, post-hoc inference time decision-rule correction, and video capsule endoscopy classification.
\subsection{Class imbalance and long-tailed recognition}
A significant portion of work addresses class imbalance via training objective: focal loss down-weights dominant classes and focuses the learning on minority classes ~\cite{lin2017focal}; class-balanced loss re-weights by the effective number of samples per class rather than raw frequency~\cite{cui2019classbalanced}; and LDAM+DRW combines a label-distribution-aware margin with deferred re-weighting~\cite{cao2019ldam}. \citet{kang2020decoupling} show that decoupling representation learning from classifier training, retraining only a lightweight classifier head under class-balanced sampling, is competitive with joint re-weighted training. All of these methods require retraining or fine-tuning the classifier, while our decoders instead operate entirely at inference time on a frozen model.

\subsection{Post-hoc and plug-in correction}
 Logit adjustment shifts the decision rule of an already-trained classifier using the label prior, either post-hoc or as a training-time loss modification, and is shown to be Bayes-consistent under label shift ~\cite{menon2021logitadjustment}. This is a special case of the broader theory of plug-in classifiers for non-decomposable performance metrics such as macro-F1, which characterizes the Bayes-optimal decision rule as a thresholded function of the posterior and a metric-dependent correction~\cite{koyejo2014consistent, narasimhan2015consistent}. Separately, optimal transport is used to frame the class imbalance problem as an approximate transport problem from training distribution to match test distribution~\cite{jin2023optransport}. The OT baseline in our experiments implements this via entropic-regularized optimal transport, solved with Sinkhorn's matrix scaling algorithm~\cite{cuturi2013sinkhorn}, which converts the classical (and computationally prohibitive) transport problem into one solvable at the scale needed for frame-level decoding. Our decoders build on this plug-in family but condition on video-level, not just global class priors, which we show is necessary to escape the base-rate floor these single-frame corrections face.

\subsection{Robust feature learning}
Separate from decision-rule correction, robust representation learning by exploring the frequency domain of images, contrastive learning to orient the model to learn highly separable discriminative features~\cite{ranabhat2025shapebias, khosla2020supcon} makes the model rely on robust features like shape. These methods which are expected to rely on robust features and reduce texture bias are also insufficient to rescue the long tailed classes.

\subsection{VCE datasets and classification}
Kvasir-Capsule~\cite{smedsrud2021kvasircapsule} consists of 117 videos and more than 47,000 labelled frames across 14 classes from real VCE examinations. GALAR is a more recent, larger, multi-center, multi-label VCE dataset having 80 videos and 3.5 million annotated frames across 29 anatomical, functional, and pathological labels~\cite{lefloch2025galar}. Both datasets have the same structure our method exploits: strong correlation between frames drawn from the same patient video. Most prior classification work treat each frame independently at inference time, whereas we use the video structure directly in the decoding rule. Aggregation of frames rather taking each frame independently is a well-known idea in video classification, from segment-level consensus on sparsely sampled snippets~\cite{wang2016tsn} to temporal smoothing of per-frame class probabilities in order to suppress single-frame noise~\cite{son2022smallbowel}. Our temporal-smoothing baseline follows this same idea, averaging posteriors within a sliding window along each video. Our contribution is the addition of a non-local conditioning on video-level disease presence that a local temporal window alone does not provide.

\subsection{Architecture baselines}
For the classifier baseline, we compared various models from the standard CNN and transformer families. DenseNet, which connects each layer to every preceding layer to encourage feature reuse~\cite{huang2017densenet}; ResNet, which uses residual connections to enable training much deeper networks~\cite{he2016resnet}; EfficientNet, which uniformly scales depth, width, and resolution via a compound coefficient~\cite{tan2019efficientnet}; Vision Transformer (ViT), pure transformer applied to patched images~\cite{dosovitskiy2021vit}; Swin Transformer, a hierarchical transformer using shifted local windows for linear-complexity self-attention~\cite{liu2021swin}; DINOv2, a self-supervised foundation model producing general-purpose visual features~\cite{oquab2023dinov2}; FocalConvNet, a focal-modulation network proposed specifically for anatomical and pathological classification on Kvasir-Capsule~\cite{srivastava2022focalconvnet}, and DenseNet201+Swin-T, a dual-stream backbone used throughout our experiments, combining a DenseNet CNN branch for local features with a Swin Transformer branch for global context, is adopted from \citet{subedi2025cnntransformer} were compared to choose a strong baseline for classifier. DenseNet201+Swin-T is used throughout our experiments as the best classifier model.

\section{Methodology}
\label{sec:methodology}
\begin{figure*}[!t]
    \centering
    \includegraphics[width=1\linewidth]{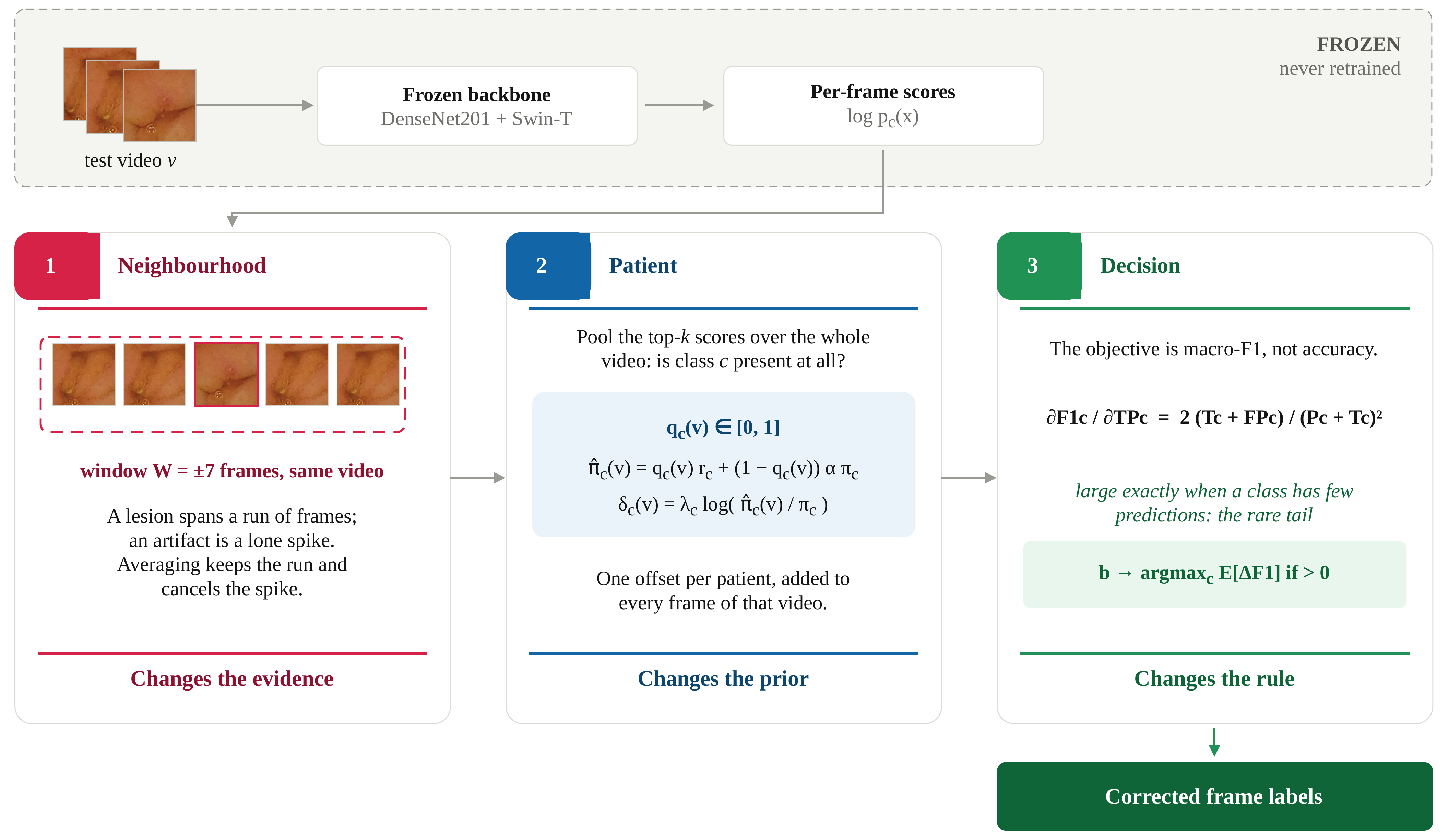}
    \caption{HCE, applied at decode time to a frozen model. The backbone (DenseNet201 and Swin-T) is never retrained; HCE reads only its per-frame log scores. Three stages run in sequence, each acting on a different term of the plug-in classifier. (1) Neighbourhood replaces each frame's scores with a mean over a $W = \pm 7$ window inside the same video, which preserves a lesion spanning a run of frames and cancels an isolated artifact spike. (2) Patient pools the top-$k$ smoothed scores over the whole video into a presence estimate $q_c(v)$, forms the video-conditioned prior $\hat{\pi}_c(v)$, and adds the single offset $\delta_c(v) = \lambda_c \log(\hat{\pi}_c(v)/\pi_c)$ to every frame of that video. (3) Decision replaces the argmax itself: a frame moves to the class maximising the expected macro-F1 gain, and only when that gain is positive. Because $\partial F1_c / \partial TP_c = 2(T_c + FP_c)/(P_c + T_c)^2$ is largest for classes with few predictions, the rule favours the rare tail. Together the three stages change the evidence, the prior and the rule respectively, which is why none of them is redundant.}
    \label{fig:hce_architecture}
\end{figure*}
\subsection{Setup}
Let a frame $x$ from video $v$ has frozen classifier posterior $p_c(x)$ for $K$ disease classes, produced by a forward pass of a trained classifier. Since, each video belongs to a single patient, the frames inside the video are highly correlated. Our decoders only modify the decision rule applied to $p_c(x)$ at inference time without the need of re-training or test-time labels. The conditional probability for the presence of a disease in the entire video is computed using the posteriors. This conditional serves as an additional global information for each local frame decoding.

\subsection{Temporal smoothing (preprocessing)} 
As a preprocessing step, we average log-posteriors within a sliding window of $W$ frames in each video. This is a standard aggregation practice for processing correlated video frames. Our temporal-smoothing baseline uses this exact method. Any video presence conditioning or gain arbitration are done after this preprocessing. HCE undergoes temporal-smoothing as a preprocessing step. Any improvement it shows is on top of temporal smoothing, not on the raw per-frame ones.


\subsection{Expected Gain Arbitration (EGA)}
EGA is a per-frame, macro-F1-optimal correction over the confusion matrix: given a frame's posterior and its base prediction, EGA reassigns the predicted class only when doing so has positive expected macro-F1 gain, computed via greedy coordinate ascent over validation-fit class statistics. EGA is our proposed greedy solver for the macro-F1 plug-in fixed point studied by ~\citet{parambath2014},\citet{koyejo2014consistent}, and \citet{narasimhan2015}, which prior work solves via Frank-Wolfe; logit adjustment~~\cite{menon2021logitadjustment} is recovered as the special case $g_c = \text{const}$ (Appendix). Since $F1_c = 2\,TP_c/(P_c+T_c)$, reassigning one frame has a closed-form expected effect on macro-F1:

\begin{equation}
\label{eq:ega_partials}
\frac{\partial F1_c}{\partial TP_c} = \frac{2(T_c + FP_c)}{(P_c + T_c)^2}
, \qquad
\frac{\partial F1_c}{\partial FP_c} = -\frac{2\,TP_c}{(P_c+T_c)^2}.
\end{equation}
Writing $g^+_c := \partial F1_c/\partial TP_c$ and $g^-_c := \partial F1_c/\partial FP_c$, both evaluated at validation-fitted $(T_c, P_c, TP_c)$ under the base decoder, moving a frame $x$ from its current class $b$ to class $c$ has expected effect
\begin{equation}
\label{eq:ega_gain}
\begin{split}
\mathbb{E}[\Delta](x, b{\to}c) = &\big[p_c(x)\,g^+_c + (1-p_c(x))\,g^-_c\big] \\
&- \big[p_b(x)\,g^+_b + (1-p_b(x))\,g^-_b\big]
\end{split}
\end{equation}

and EGA reassigns $x$ to the best $c$ only when this is positive, iterated on validation so the fitted counts reflect earlier reassignments; at test time each frame is scored in a single pass against those frozen statistics. Eqs.~(\ref{eq:ega_partials})--(\ref{eq:ega_gain}) give the first-order form; the implementation uses the exact finite difference of $F_{\beta_c}$, with $\beta_c = 4$ on the bottom third of classes by validation prior and $1$ elsewhere, so $\beta_c = 1$ throughout recovers macro-F1. EGA depends only on frame posteriors and validation-fit statistics, it does not require the video-level presence conditioning of Section~3.4, so it can be used standalone (Tables~\ref{tab:main_kvasir} and~\ref{tab:main_galar_pathology}) or composed with conditioning as in HCE.

\subsection{Video-level presence conditioning}
After temporal smoothing, the video-level presence conditioning is computed for each class. For each class $c$ in a video $v$, the posteriors across all frames are sorted and the top-$k$ aggregate is taken and passed through a logistic model fit by cross-fitting which is the presence probability $q_c(v)$.

We then form a video-conditioned effective prior,
\begin{equation}
\hat\pi_c(v) = q_c(v) \cdot r_c + \bigl(1 - q_c(v)\bigr) \cdot \alpha \cdot \pi_c
\label{eq:effective_prior}
\end{equation}
where $\pi_c$ is the global class prior, $r_c$ is the typical within-video frame rate of class $c$ given the video contains it, and $\alpha \in [0,1]$ shrinks the prior toward zero when the video looks negative for $c$.

This is applied as an additive log-ratio offset to the frame's log-posterior,
\begin{equation}
\delta_c(v) = \lambda_c \cdot \log!\frac{\hat\pi_c(v)}{\pi_c}
\label{eq:offset}
\end{equation}
scaled by a per-class reliability weight
\begin{equation}
\lambda_c = \lambda \cdot 2\max\bigl(\mathrm{AUC}_c - 0.5,\ 0\bigr)
\label{eq:reliability}
\end{equation}
the cross-fitted AUC of $q_c(\cdot)$ against true per-video presence.

This weighting in Eq.\eqref{eq:reliability} is what keeps the method honest: for a class whose presence cannot be predicted better than chance, $\lambda_c \to 0$ and the offset in Eq.\eqref{eq:offset} vanishes entirely, rather than applying a large, unreliable boost to exactly the rarest, hardest-to-verify classes. Classes with no validation support receive no offset. We found conditioning at finer sub-video granularity than the whole video (applying Eq.~\eqref{eq:effective_prior} per chunk rather than per video) gives no consistent improvement (Appendix), so we condition at the whole-video level throughout.

\subsection{Hierarchical Conditional-Expected-gain}
HCE combines the video conditioning offset from Section 3.4 with EGA arbitration (Section 3.3). It first computes the adjusted log-posterior
\begin{equation}
\tilde{\ell}_c(x) = \log p_c(x) + \delta_c(v)
\label{eq:hce_adjust}
\end{equation}
for frame $x$ in video $v$, then applies EGA arbitration to $\mathrm{softmax}\bigl(\tilde{\ell}(x)\bigr)$ rather than to the raw posterior $p(x)$.

The two mechanisms are orthogonal, Eq.\eqref{eq:offset} shifts the effective class prior per video, EGA arbitrates globally over the confusion matrix, and each recovers cases the other misses: when the reliability weighting in Eq.\eqref{eq:reliability} drives $\lambda_c \to 0$ for a class, HCE's fallback in Eq.~\eqref{eq:hce_adjust} is EGA's global correction rather than uncorrected temporal smoothing, which is what makes the composition outperform either mechanism alone .

\section{Datasets \& Experiments}
\label{sec:datasets}
We evaluate on two independent VCE datasets. Kvasir-Capsule ~\cite{smedsrud2021kvasircapsule}: contains 117 videos with 47,238 labeled frames that come from 43 videos and has an official two-fold split. There are 14 classes, where 11 classes are practically usable as the 3 dropped classes, each have only 1 patient data which may only go in one of the folds causing 0 support in the other fold. Rare classes (bottom third by prevalence): Erosion (1.06\%), Blood-fresh (0.95\%), Erythema (0.34\%). GALAR ~\cite{lefloch2025galar} is another large recent VCE dataset that consists of 80 videos (29 classes). 12-class video-disjoint pathology subset was created keeping the single label frames divided into 2 folds. Rare classes: Ulcer (0.17\%), Erythema (0.07\%), Foreign body (0.0017\%), IBD (0.0004\%). 



DenseNet201+Swin-T was trained with the same recipe across all configurations: AdamW, learning rate $10^{-4}$, weight decay 0.01, cosine schedule, batch size 32, $224\times224$ input resolution on Tesla-V100-SXM2-32GB GPU. Kvasir-Capsule was trained upto 100 epochs (best model based on macro-F1). While GALAR was trained for 30 epochs with early stopping patience of 8. We used batch size of 64, mixed-precision training, and majority classes were capped at 40,000 training frames each for tractability, with validation subsampled to 10\% during training for monitoring only (label-blind). All logits used for decoding, and every reported result, use the complete validation and test splits. We selected DenseNet201+Swin-T  ~\cite{subedi2025cnntransformer} with focal loss as the classifier to carryout all the decoding experiments based on a series of experiments (Appendix). 2 folds $\times$ 3 seeds per dataset, were evaluated both single-model and as a 3-seed ensemble. Results are mean $\pm$ sd (over seeds for single-model, over folds for ensemble). Metrics reported are macro-F1, macro-recall, MCC, rare-class recall/precision. All decoder hyperparameters are selected using leave-video-out cross-validation without test data leakage and keeping it fully inductive. Significance test uses a per-video cluster bootstrap with a practical-significance floor (0.005 for F1/MCC, 0.003 for rare recall) to prevent a negligible but technically significant effect from 
looking meaningful.


\section{Results and Discussion}
\label{sec:results}

\begin{figure*}
    \centering
    \includegraphics[width=0.99\linewidth]{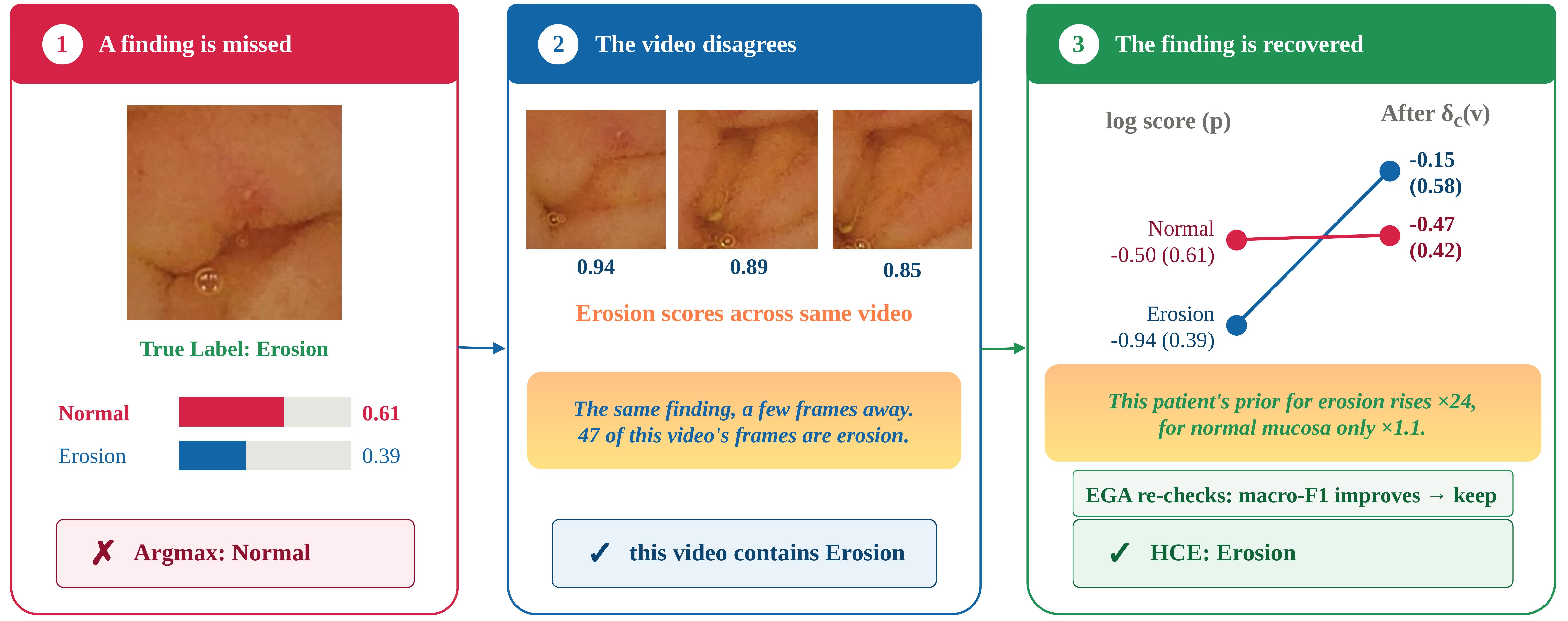}
    \caption{How a missed rare finding is recovered. A single Kvasir-Capsule frame (video eb0203196e284797, frame 6256, fold 0, seed 42) whose true label is Erosion. (1) The frozen backbone scores it Normal 0.61 against Erosion 0.39, so argmax discards it. (2) The rest of the same video disagrees: 47 of its frames carry erosion, and top-$k$ pooling turns that into a per-video prior that rises $\times 24$ for erosion but only $\times 1.1$ for normal mucosa. (3) The offset $\delta_c(v)$ added to every frame of the video reverses the ordering to Erosion 0.58 against Normal 0.42, and the expected-gain arbitration keeps the reassignment because it raises macro-F1. Log scores carry their probability in brackets; the right-hand column is renormalised, since adding $\delta$ breaks normalisation. Both stages always run: when a video is uninformative the gate sets $\delta_c(v) = 0$ and arbitration recovers the finding on its own.}
    \label{fig:rescue_example}
\end{figure*}

\begin{table*}[t]
  \centering
  \small
  \begin{tabular}{lccccc}
    \toprule
    Method & Macro-F1 & Macro Recall & MCC & Rare Recall & Rare Precision \\
    \midrule
    \multicolumn{6}{l}{\textit{Single model}} \\
    \midrule
    Argmax (backbone)~\cite{subedi2025cnntransformer} & 0.2403$\pm$0.0274 & 0.2664$\pm$0.0163 & 0.3683$\pm$0.0374 & 0.0511$\pm$0.0264 & 0.0307$\pm$0.0199 \\
    LA~\cite{menon2021logitadjustment} & 0.2403$\pm$0.0274 & 0.2664$\pm$0.0163 & 0.3683$\pm$0.0374 & 0.0511$\pm$0.0264 & 0.0307$\pm$0.0199 \\
    OT~\cite{jin2023optransport} & 0.2489$\pm$0.0271 & 0.2954$\pm$0.0225 & 0.3891$\pm$0.0301 & 0.0477$\pm$0.0262 & \textbf{0.0413$\pm$0.0380} \\
    Temporal (baseline)~\cite{son2022smallbowel} & 0.2435$\pm$0.0276 & 0.2656$\pm$0.0178 & 0.3813$\pm$0.0385 & 0.0372$\pm$0.0266 & 0.0213$\pm$0.0189 \\
    EGA (ours) & 0.2533$\pm$0.0301 & 0.2730$\pm$0.0148 & 0.3774$\pm$0.0462 & \textbf{0.0700$\pm$0.0561} & 0.0393$\pm$0.0382 \\
    \textbf{HCE (ours)} & \textbf{0.2654$\pm$0.0413} & \textbf{0.2986$\pm$0.0311} & \textbf{0.3955$\pm$0.0573} & 0.0635$\pm$0.0345 & 0.0191$\pm$0.0130 \\
    \midrule
    \multicolumn{6}{l}{\textit{3-seed ensemble}} \\
    \midrule
    Argmax & 0.2379$\pm$0.0231 & 0.2604$\pm$0.0053 & 0.3753$\pm$0.0179 & 0.0399$\pm$0.0254 & 0.0186$\pm$0.0108 \\
    LA & 0.2545$\pm$0.0233 & 0.2889$\pm$0.0068 & 0.3925$\pm$0.0345 & 0.0635$\pm$0.0432 & 0.0195$\pm$0.0097 \\
    OT & 0.2627$\pm$0.0290 & 0.3463$\pm$0.0151 & 0.4077$\pm$0.0425 & 0.0820$\pm$0.0724 & \textbf{0.0255$\pm$0.0136} \\
    Temporal (baseline) & 0.2357$\pm$0.0169 & 0.2573$\pm$0.0083 & 0.3795$\pm$0.0086 & 0.0296$\pm$0.0161 & 0.0174$\pm$0.0118 \\
    EGA (ours) & 0.2779$\pm$0.0273 & 0.3046$\pm$0.0006 & 0.4174$\pm$0.0398 & 0.0707$\pm$0.0426 & 0.0195$\pm$0.0089 \\
    \textbf{HCE (ours)} & \textbf{0.3023$\pm$0.0243} & \textbf{0.3472$\pm$0.0047} & \textbf{0.4421$\pm$0.0569} & \textbf{0.1013$\pm$0.0578} & 0.0224$\pm$0.0112 \\
    \bottomrule
  \end{tabular}
  \caption{Kvasir-Capsule, decoding methods. Mean$\pm$sd over 3 seeds $\times$ 2 folds (single) / 2 folds (ensemble). \textbf{Bold} = best per column (within this block). $\uparrow$/$\downarrow$: bootstrap 95\% CI excludes zero (per-video cluster bootstrap) AND practically meaningful ($|\Delta|\geq$0.003-0.005) vs. temporal baseline. Unmarked = CI includes zero or delta below the practical floor.}
  \label{tab:main_kvasir}
\end{table*}

\begin{table*}[t]
  \centering
  \small
  \begin{tabular}{lccccc}
    \toprule
    Method & Macro-F1 & Macro Recall & MCC & Rare Recall & Rare Precision \\
    \midrule
    \multicolumn{6}{l}{\textit{Single model}} \\
    \midrule
    Argmax & 0.1839$\pm$0.0159 & 0.1990$\pm$0.0226 & 0.4195$\pm$0.0824 & 0.0250$\pm$0.0180 & 0.0618$\pm$0.0573 \\
    LA & 0.1839$\pm$0.0159 & 0.1990$\pm$0.0226 & 0.4195$\pm$0.0824 & 0.0250$\pm$0.0180 & 0.0618$\pm$0.0573 \\
    OT & 0.1671$\pm$0.0156 & 0.1884$\pm$0.0416 & 0.3479$\pm$0.0411 & 0.0304$\pm$0.0290 & 0.0250$\pm$0.0255 \\
    Temporal (baseline) & 0.1878$\pm$0.0156 & 0.1964$\pm$0.0202 & 0.4486$\pm$0.0851 & 0.0219$\pm$0.0165 & \textbf{0.0704$\pm$0.0828} \\
    EGA (ours) & \textbf{0.1924$\pm$0.0194} & 0.2008$\pm$0.0252 & \textbf{0.4516$\pm$0.0828} & 0.0445$\pm$0.0308 & 0.0568$\pm$0.0532 \\
    \textbf{HCE (ours)} & 0.1894$\pm$0.0187 & \textbf{0.2052$\pm$0.0231} & 0.4421$\pm$0.0794 & \textbf{0.0590$\pm$0.0361$^{\uparrow}$} & 0.0487$\pm$0.0531 \\
    \midrule
    \multicolumn{6}{l}{\textit{3-seed ensemble}} \\
    \midrule
    Argmax & 0.1947$\pm$0.0147 & 0.2008$\pm$0.0199 & 0.4612$\pm$0.0921 & 0.0219$\pm$0.0139 & 0.0846$\pm$0.0763 \\
    LA & 0.1947$\pm$0.0147 & 0.2008$\pm$0.0199 & 0.4612$\pm$0.0921 & 0.0219$\pm$0.0139 & 0.0846$\pm$0.0763 \\
    OT & 0.1755$\pm$0.0196 & 0.1986$\pm$0.0472 & 0.3630$\pm$0.0207 & 0.0436$\pm$0.0360 & 0.0411$\pm$0.0358 \\
    Temporal (baseline) & 0.1979$\pm$0.0148 & 0.1978$\pm$0.0189 & 0.4798$\pm$0.0929 & 0.0193$\pm$0.0137 & \textbf{0.1222$\pm$0.1123} \\
    EGA (ours) & \textbf{0.2032$\pm$0.0190} & 0.2076$\pm$0.0283 & \textbf{0.4819$\pm$0.0911} & 0.0583$\pm$0.0429 & 0.0760$\pm$0.0686 \\
    \textbf{HCE (ours)} & 0.2000$\pm$0.0170 & \textbf{0.2092$\pm$0.0268} & 0.4772$\pm$0.0911 & \textbf{0.0598$\pm$0.0404$^{\uparrow}$} & 0.0798$\pm$0.0727 \\
    \bottomrule
  \end{tabular}
  \caption{GALAR, decoding methods. Mean$\pm$SD over 3 seeds $\times$ 2 folds (single) / 2 folds (ensemble). \textbf{Bold} = best per column (within this block). $\uparrow$/$\downarrow$: bootstrap 95\% CI excludes zero (per-video cluster bootstrap) AND practically meaningful ($|\Delta|\geq$0.003-0.005) vs. temporal baseline. Unmarked = CI includes zero or delta below the practical floor.}
  \label{tab:main_galar_pathology}
\end{table*}

\begin{figure*}[!t]
    \centering
    \includegraphics[width=0.99\linewidth]{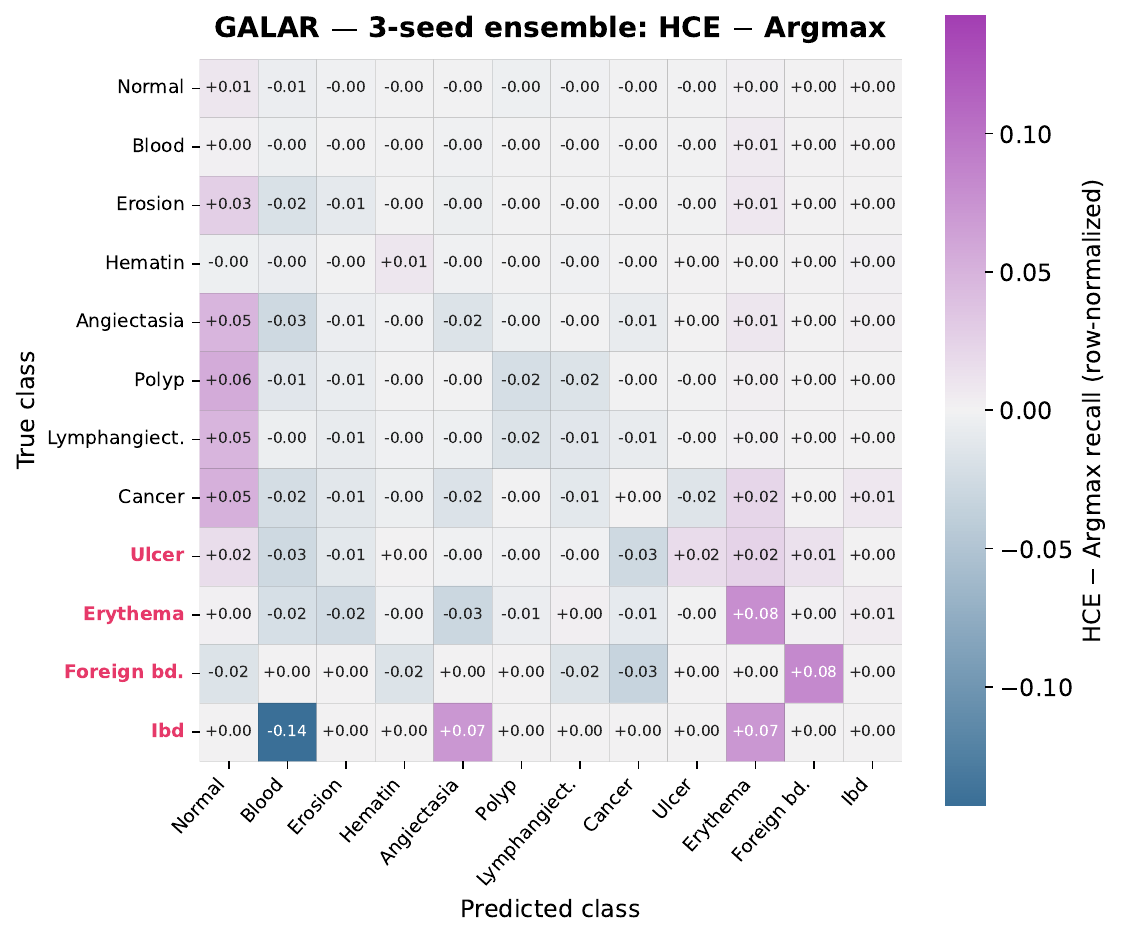}
    \caption{HCE minus Argmax, row-normalized confusion matrix delta, GALAR ensemble. Diagonal values are the recall change per class. \textcolor{rarepink}{Erythema, foreign body, and ulcer} gain recall (+0.079, +0.083, +0.017). IBD is flat (14 test frames). Polyp, angiectasia, lymphangioectasis, and erosion lose recall (-0.022, -0.016, -0.013, -0.010), most of that loss lands on ``normal''. Erythema and ulcer gains come from confusion with other lesion and hemorrhage classes.}
    \label{fig:confusion_rescue}
\end{figure*}


Tables~\ref{tab:main_kvasir} and \ref{tab:main_galar_pathology} summarize decoding-method performance across both datasets, single-model and 3-seed ensemble. HCE gives the best macro recall of any method in every evaluation setting, and the best rare-class recall in three of the four, EGA edges ahead on Kvasir-Capsule single-model (0.0700 vs.\ 0.0635) but HCE leads on Kvasir-Capsule ensemble and both GALAR scopes. Against the temporal-smoothing baseline, this rare-recall gain is 2.7--3.1x on GALAR (+0.036, 95\% CI[0.010,0.068], single; +0.037, CI[0.005,0.067], ensemble) and up to 3.4x on Kvasir-Capsule ensemble (+0.072, CI[0.000,0.180]). Against OT, HCE's rare-class recall gain is +0.016 (Kvasir-single, CI[-0.033,0.114]), +0.019 (Kvasir-ensemble, CI[-0.077,0.195]), +0.029 (GALAR-single, CI[-0.009,0.071]), and +0.016 (GALAR-ensemble, CI[-0.011,0.045]); all four CIs include zero, so none reach significance at this sample size. HCE does significantly improve macro-F1 over OT on GALAR (+0.022, CI[0.006,0.036], single; +0.025, CI[0.003,0.044], ensemble) and MCC on GALAR-single (+0.094, CI[0.004,0.168]); the GALAR-ensemble MCC gain is large in point estimate (+0.114) but its CI narrowly includes zero (CI[-0.001,0.213]). On Kvasir-Capsule, HCE also achieves the best macro-F1 and MCC outright; on GALAR, EGA's plain per-frame arbitration alone edges out HCE on F1 and MCC (by 0.003--0.01), reflecting a real trade, video-level context composed with EGA is what drives the large rare-recall gain there, at a small cost to overall balance. HCE's own tradeoff is rare-class precision, where OT (Kvasir) and temporal (GALAR) both stay ahead, consistent with a method that intentionally shifts decisions toward under-detected classes. LA reduces to argmax in three of the four blocks: the validation sweep over $\tau$ selects $\tau = 0$, so no adjustment is macro-F1-optimal there.

As shown in the confusion matrix for recall rescue in~\ref{fig:confusion_rescue}, GALAR's 3-seed ensemble, HCE raises recall on three of the four rarest classes relative to argmax: erythema (0.076→0.156, +0.079), foreign body (0.000→0.083, +0.083, rescued from total miss), and ulcer (0.011→0.028, +0.017). IBD stays at zero recall under both methods, its test support is 14 frames, too little signal for any decoder to act on. These gains are paid for mainly by four mid-frequency lesion classes: polyp (-0.022), angiectasia (-0.016), lymphangioectasis (-0.013), and erosion (-0.010). The majority "normal" class is not degraded, its own recall rises slightly (+0.010). We can see that the rare class recall gain is large indicated by large positive values, while some pathologies were missed to normal as shown in the mid section of the normal-predicted column.

\begin{table*}[t]
  \centering
  \small
  \begin{tabular}{lcccc|ccc}
    \toprule
    & \multicolumn{4}{c|}{Rare Recall} & \multicolumn{3}{c}{Mean over settings} \\
    Variant & Kv-S & Kv-E & GA-S & GA-E & Rare Rec. & F1 & MCC \\
    \midrule
    Temporal & 0.037 & 0.030 & 0.022 & 0.019 & 0.027 & 0.216 & 0.422 \\
    Cond.\ (video) & 0.047 & 0.074 & 0.025 & 0.024 & 0.042 & 0.225 & 0.430 \\
    Cond.\ (+rel.\ wt.) & 0.049 & 0.066 & 0.025 & 0.022 & 0.040 & 0.227 & \textbf{0.438} \\
    Cond.\ (budgeted) & 0.043 & 0.052 & 0.023 & 0.021 & 0.035 & 0.223 & 0.435 \\
    Cond.\ (asym.) & 0.049 & 0.066 & 0.025 & 0.022 & 0.040 & 0.227 & \textbf{0.439} \\
    EGA alone & \textbf{0.070} & 0.071 & 0.044 & \textbf{0.058} & 0.061 & 0.232 & 0.432 \\
    Auto-select$^{\dagger}$ & 0.049 & 0.073 & 0.043 & 0.055 & 0.055 & 0.231 & 0.436 \\

    PACT (constrained) & 0.048 & 0.065 & 0.030 & 0.054 & 0.049 & 0.226 & 0.431 \\

    \textbf{HCE (composed)} & 0.064 & \textbf{0.101} & \textbf{0.059} & \textbf{0.060} & \textbf{0.071} & \textbf{0.239} & \textbf{0.439} \\
    \bottomrule
  \end{tabular}
  \caption{Ablation: conditioning mechanism, all four settings (Kv/GA = Kvasir-Capsule/GALAR, S/E = single/3-seed ensemble). \textbf{Bold} = best per column, ties within 0.0015 bolded jointly. The datasets separate the mechanism: on Kvasir, conditioning alone lifts rare recall over Temporal; on GALAR it is inert, consistent with its videos containing most classes, leaving the video-level presence estimate near-constant. Composition exceeds both components on 3/4 settings; on Kv-S, EGA alone edges it out on rare recall. Means are unweighted over the four settings, with no sd, as the spread is between-dataset prevalence rather than measurement noise. $^{\dagger}$hce\_auto selects among the three on validation (selection leak); shown for completeness, excluded from bolding, not a reported result. \textbf{PACT}=Presence-Aware Constrained Tuning
}
  \label{tab:ablation_mechanism}
\end{table*}

\begin{figure*}[t]
    \centering
    \includegraphics[width=0.99\linewidth]{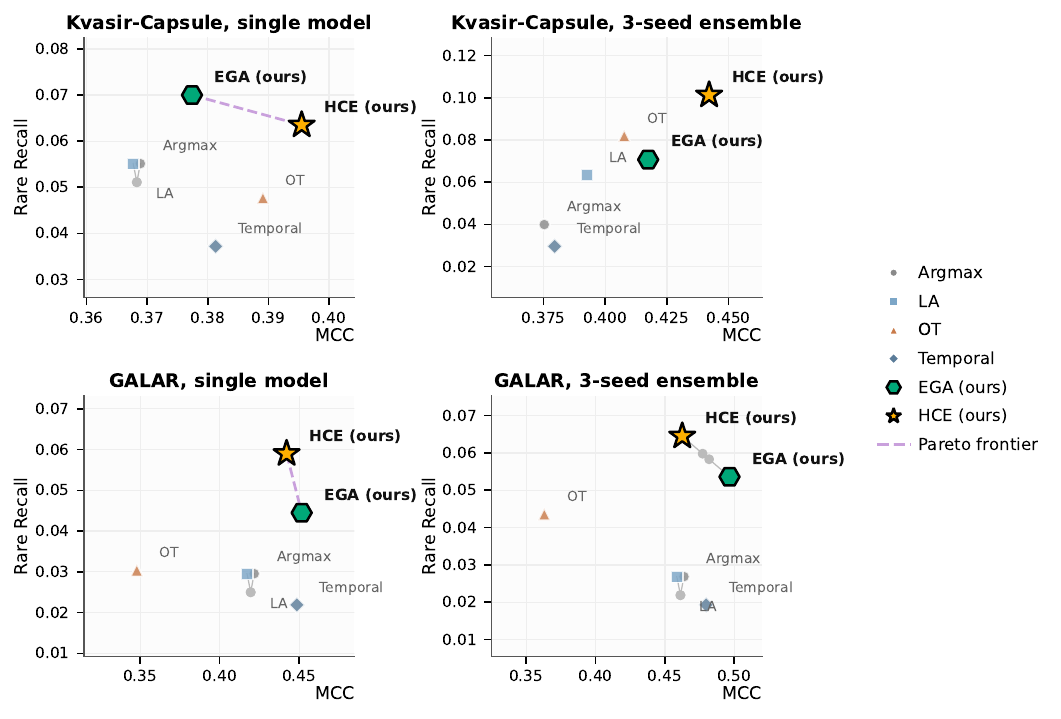}
    \caption{MCC vs. rare recall for every decoding method, one panel per dataset and scope (Kvasir-Capsule/GALAR, single model/3-seed ensemble). Baselines (argmax, LA, OT, temporal) are small and muted; EGA and HCE are shown bold. HCE sits at top right in almost all settings except flat MCC on GALAR and rare-recall slightly less than EGA on Kvasir-Capsule single.}
    \label{fig:pareto_methods}
\end{figure*}

\subsection{Ablations}
\label{sec:ablations}

\ref{tab:ablation_mechanism} reports the mechanism ablation on all four evaluation settings, since the two datasets separate the components differently. Video-level conditioning alone, in any of its four variants (plain, reliability-weighted, budgeted, asymmetric), barely moves rare recall on GALAR (0.021 to 0.025 against a temporal baseline of 0.019 to 0.022), but on Kvasir-Capsule it produces a real gain by itself (0.047 single, 0.074 ensemble, against 0.037 and 0.030). This is the presence-variance argument showing up in data: GALAR videos contain most classes, so $q_c(v)$ barely varies and the offset degenerates toward a global one, while Kvasir- Capsule videos do not. EGA's per-frame arbitration is the stronger single component on average (mean 0.061 vs 0.042). Composing the two gives the best mean rare recall (0.071) and the best result in three of the four settings, including the largest gain of the table on Kvasir-Capsule ensemble (0.101, against 0.074 for conditioning alone and 0.071 for EGA alone). The exception is Kvasir-Capsule single-model, where EGA alone edges ahead (0.070 vs 0.064). The composition's cost is GALAR balance: against EGA alone it gives up macro-F1 and MCC there (0.189/0.442 vs 0.192/0.452 single, 0.200/0.477 vs 0.203/0.482 ensemble), though averaged over the four settings the two are indistinguishable on MCC and HCE leads on macro-F1. The MCC-constrained composition protects balance but returns most of the rare-recall gain (mean 0.049 against 0.071).

\subsection{Discussion \& Analysis }
\label{sec:disc_analysis}

The MCC vs rare-recall Pareto plot shown in ~\ref{fig:pareto_methods} clearly shows our methods EGA and HCE always placed at the upper-right corner, while the baselines (argmax, LA, OT, temporal) cluster in a visibly distinguishable lower-left region. HCE wins on rare recall across all settings except for Kvasir-Capsule single model. While on MCC, it clearly wins on Kvasir-Capsule, but loses slightly on GALAR to EGA and temporal baslines, but it is comparable to other baselines like argmax and LA. OT MCC collapses on GALAR but wins other baselines on Kvasir-Capsule. The OT rare recall is slightly better than baselines on ensemble while comparable to baselines on single models. Critically, on Kvasir-Capsule HCE does not just sit on the frontier, it dominates every baseline outright, better MCC and better rare recall simultaneously, showing the gain is not merely a recall/precision-style trade sliding along an existing frontier, but a genuine expansion of the achievable region: video-level conditioning composed with EGA's arbitration reaches a point no single-frame, post-hoc correction (LA, OT) or purely temporal baseline can reach on either axis alone.

\subsection{Limitations \& Future Work}
\label{sec:limit_future}

Significance testing is underpowered (2 CV folds), so most gains are directionally consistent, but not statistically confirmed except GALAR rare recall. Reliability weighting sets $\lambda_c \to 0$ when a class's cross-fitted presence AUC is near chance, which is the regime for the very rarest classes. Video-level conditioning is therefore inactive precisely where rare-class support is thinnest, and HCE degenerates to EGA on that tail. The rare recall came with a rare precision cost and generalization beyond capsule endoscopy is untested. 

Reducing rare class false-positive rate via a confirmation stage before committing a rare-class label could be a natural next thing to try. Extending HCE mechanism to non-endoscopy imbalanced video domain can reveal the cross-domain usefulness of this method. Two baselines we did not run would sharpen the comparison: a Frank-Wolfe solver for the same macro-F1 plug-in fixed point EGA targets, which would isolate the value of our greedy solver specifically, and per-class threshold tuning on validation as a simpler post-hoc alternative.

\section{Conclusion}
\label{sec:conclusion}

We presented HCE, a training-free, video-conditioned decoding method that composes a macro-F1-optimal per-frame correction (EGA) with video-level disease-presence conditioning to improve rare-disease recall in capsule endoscopy triage, without retraining the classifier or using test-time labels. Across two datasets, HCE improves rare-class recall over both a temporal-smoothing baseline and an optimal-transport correction in all four evaluation settings. On GALAR, the gain over temporal smoothing is statistically confirmed in both scopes, as are gains over optimal transport in macro F1 and, in the single-model setting, MCC. Comparisons on Kvasir-Capsule and rare-recall comparisons against optimal transport are directionally consistent, but their intervals include zero. A precision trade-off on rare classes and a small F1/MCC drop relative to EGA alone on GALAR was observed. These results support HCE as a practical, low-cost way to surface rare-disease frames within a capsule study, not as an autonomous diagnostic decision.

\bibliographystyle{plainnat}
\bibliography{main}

@String(CVPR  = {IEEE Conf. Comput. Vis. Pattern Recog.})

@String(ICCV  = {Int. Conf. Comput. Vis.})

@String(CVPR  = {CVPR})

@String(ICCV  = {ICCV})

@inproceedings{lin2017focal,
  title={Focal loss for dense object detection},
  author={Lin, Tsung-Yi and Goyal, Priya and Girshick, Ross and He, Kaiming and Doll{\'a}r, Piotr},
  booktitle={Proceedings of the IEEE international conference on computer vision},
  pages={2980--2988},
  year={2017}
}

@inproceedings{cui2019classbalanced,
  title={Class-balanced loss based on effective number of samples},
  author={Cui, Yin and Jia, Menglin and Lin, Tsung-Yi and Song, Yang and Belongie, Serge},
  booktitle={2019 IEEE/CVF conference on computer vision and pattern recognition (CVPR)},
  pages={9260--9269},
  year={2019},
  organization={IEEE}
}

@article{cao2019ldam,
  title={Learning imbalanced datasets with label-distribution-aware margin loss},
  author={Cao, Kaidi and Wei, Colin and Gaidon, Adrien and Arechiga, Nikos and Ma, Tengyu},
  journal={Advances in neural information processing systems},
  volume={32},
  year={2019}
}

@article{kang2020decoupling,
  title={Decoupling representation and classifier for long-tailed recognition},
  author={Kang, Bingyi and Xie, Saining and Rohrbach, Marcus and Yan, Zhicheng and Gordo, Albert and Feng, Jiashi and Kalantidis, Yannis},
  journal={arXiv preprint arXiv:1910.09217},
  year={2019}
}

@article{menon2021logitadjustment,
  title={Long-tail learning via logit adjustment},
  author={Menon, Aditya Krishna and Jayasumana, Sadeep and Rawat, Ankit Singh and Jain, Himanshu and Veit, Andreas and Kumar, Sanjiv},
  journal={arXiv preprint arXiv:2007.07314},
  year={2020}
}

@article{koyejo2014consistent,
  title={Consistent binary classification with generalized performance metrics},
  author={Koyejo, Oluwasanmi and Natarajan, Nagarajan and Ravikumar, Pradeep and Dhillon, Inderjit S},
  journal={Advances in neural information processing systems},
  volume={27},
  year={2014}
}

@article{narasimhan2015consistent,
  title={On the statistical consistency of plug-in classifiers for non-decomposable performance measures},
  author={Narasimhan, Harikrishna and Vaish, Rohit and Agarwal, Shivani},
  journal={Advances in neural information processing systems},
  volume={27},
  year={2014}
}

@article{jin2023optransport,
  title={An optimal transport view of class-imbalanced visual recognition},
  author={Jin, Lianbao and Lang, Dayu and Lei, Na},
  journal={International Journal of Computer Vision},
  volume={131},
  number={11},
  pages={2845--2863},
  year={2023},
  publisher={Springer}
}

@article{smedsrud2021kvasircapsule,
  title={Kvasir-Capsule, a video capsule endoscopy dataset},
  author={Smedsrud, Pia H and Thambawita, Vajira and Hicks, Steven A and Gjestang, Henrik and Nedrejord, Oda Olsen and N{\ae}ss, Espen and Borgli, Hanna and Jha, Debesh and Berstad, Tor Jan Derek and Eskeland, Sigrun L and others},
  journal={Scientific Data},
  volume={8},
  number={1},
  pages={142},
  year={2021},
  publisher={Nature Publishing Group UK London}
}

@article{lefloch2025galar,
  title={Galar-a large multi-label video capsule endoscopy dataset},
  author={Le Floch, Maxime and Wolf, Fabian and McIntyre, Lucian and Weinert, Christoph and Palm, Albrecht and Volk, Konrad and Herzog, Paul and Kirk, Sophie Helene and Steinh{\"a}user, Jonas L and Stopp, Catrein and others},
  journal={Scientific Data},
  volume={12},
  number={1},
  pages={828},
  year={2025}
}

@inproceedings{ranabhat2025shapebias,
  title={Promoting shape bias in cnns: Frequency-based and contrastive regularization for corruption robustness},
  author={Ranabhat, Robin Narsingh and Wang, Longwei and Patel, Amit Kumar and Santosh, KC},
  booktitle={International Conference on Intelligent Systems and Pattern Recognition},
  pages={16--26},
  year={2025},
  organization={Springer}
}

@article{khosla2020supcon,
  title={Supervised contrastive learning},
  author={Khosla, Prannay and Teterwak, Piotr and Wang, Chen and Sarna, Aaron and Tian, Yonglong and Isola, Phillip and Maschinot, Aaron and Liu, Ce and Krishnan, Dilip},
  journal={Advances in neural information processing systems},
  volume={33},
  pages={18661--18673},
  year={2020}
}

@article{cuturi2013sinkhorn,
  title={Sinkhorn distances: Lightspeed computation of optimal transport},
  author={Cuturi, Marco},
  journal={Advances in neural information processing systems},
  volume={26},
  year={2013}
}

@inproceedings{wang2016tsn,
  title={Temporal segment networks: Towards good practices for deep action recognition},
  author={Wang, Limin and Xiong, Yuanjun and Wang, Zhe and Qiao, Yu and Lin, Dahua and Tang, Xiaoou and Van Gool, Luc},
  booktitle={European conference on computer vision},
  pages={20--36},
  year={2016},
  organization={Springer}
}

@article{son2022smallbowel,
  title={Small bowel detection for wireless capsule endoscopy using convolutional neural networks with temporal filtering},
  author={Son, Geonhui and Eo, Taejoon and An, Jiwoong and Oh, Dong Jun and Shin, Yejee and Rha, Hyenogseop and Kim, You Jin and Lim, Yun Jeong and Hwang, Dosik},
  journal={Diagnostics},
  volume={12},
  number={8},
  pages={1858},
  year={2022},
  publisher={MDPI}
}

@inproceedings{subedi2025cnntransformer,
  title={Classification of endoscopy and video capsule images using CNN-transformer model},
  author={Subedi, Aliza and Regmi, Smriti and Regmi, Nisha and Bhusal, Bhumi and Bagci, Ulas and Jha, Debesh},
  booktitle={MICCAI Workshop on Cancer Prevention through Early Detection},
  pages={26--36},
  year={2024},
  organization={Springer}
}

@inproceedings{huang2017densenet,
  title={Densely connected convolutional networks},
  author={Huang, Gao and Liu, Zhuang and Van Der Maaten, Laurens and Weinberger, Kilian Q},
  booktitle={2017 IEEE conference on computer vision and pattern recognition (CVPR)},
  pages={2261--2269},
  year={2017},
  organization={Ieee}
}

@inproceedings{he2016resnet,
  title={Deep residual learning for image recognition},
  author={He, Kaiming and Zhang, Xiangyu and Ren, Shaoqing and Sun, Jian},
  booktitle={Proceedings of the IEEE conference on computer vision and pattern recognition},
  pages={770--778},
  year={2016}
}

@inproceedings{tan2019efficientnet,
  title={Efficientnet: Rethinking model scaling for convolutional neural networks},
  author={Tan, Mingxing and Le, Quoc},
  booktitle={International conference on machine learning},
  pages={6105--6114},
  year={2019},
  organization={PmLR}
}

@article{dosovitskiy2021vit,
  title={An image is worth 16x16 words: Transformers for image recognition at scale},
  author={Dosovitskiy, Alexey and Beyer, Lucas and Kolesnikov, Alexander and Weissenborn, Dirk and Zhai, Xiaohua and Unterthiner, Thomas and Dehghani, Mostafa and Minderer, Matthias and Heigold, Georg and Gelly, Sylvain and others},
  journal={arXiv preprint arXiv:2010.11929},
  year={2020}
}

@inproceedings{liu2021swin,
  title={Swin transformer: Hierarchical vision transformer using shifted windows},
  author={Liu, Ze and Lin, Yutong and Cao, Yue and Hu, Han and Wei, Yixuan and Zhang, Zheng and Lin, Stephen and Guo, Baining},
  booktitle={2021 IEEE/CVF international conference on computer vision (ICCV)},
  pages={9992--10002},
  year={2021},
  organization={Ieee}
}

@article{oquab2023dinov2,
  title={Dinov2: Learning robust visual features without supervision},
  author={Oquab, Maxime and Darcet, Timoth{\'e}e and Moutakanni, Th{\'e}o and Vo, Huy and Szafraniec, Marc and Khalidov, Vasil and Fernandez, Pierre and Haziza, Daniel and Massa, Francisco and El-Nouby, Alaaeldin and others},
  journal={arXiv preprint arXiv:2304.07193},
  year={2023}
}

@inproceedings{srivastava2022focalconvnet,
  title={Video capsule endoscopy classification using focal modulation guided convolutional neural network},
  author={Srivastava, Abhishek and Tomar, Nikhil Kumar and Bagci, Ulas and Jha, Debesh},
  booktitle={2022 IEEE 35th international symposium on computer-based medical systems (CBMS)},
  pages={323--328},
  year={2022},
  organization={IEEE}
}

@article{parambath2014,
  title={Optimizing F-measures by cost-sensitive classification},
  author={Parambath, Shameem A and Usunier, Nicolas and Grandvalet, Yves},
  journal={Advances in neural information processing systems},
  volume={27},
  year={2014}
}

@inproceedings{narasimhan2015,
  title={Consistent multiclass algorithms for complex performance measures},
  author={Narasimhan, Harikrishna and Ramaswamy, Harish and Saha, Aadirupa and Agarwal, Shivani},
  booktitle={International Conference on Machine Learning},
  pages={2398--2407},
  year={2015},
  organization={PMLR}
}

@article{iddan2000wireless,
  title={Wireless capsule endoscopy},
  author={Iddan, Gavriel and Meron, Gavriel and Glukhovsky, Arkady and Swain, Paul},
  journal={Nature},
  volume={405},
  number={6785},
  pages={417--417},
  year={2000},
  publisher={Nature Publishing Group UK London}
}

@article{zheng2012detection,
  title={Detection of lesions during capsule endoscopy: physician performance is disappointing},
  author={Zheng, YuanPu and Hawkins, Lauren and Wolff, Jordan and Goloubeva, Olga and Goldberg, Eric},
  journal={Official journal of the American College of Gastroenterology| ACG},
  volume={107},
  number={4},
  pages={554--560},
  year={2012},
  publisher={LWW}
}

@article{buda2018systematic,
  title={A systematic study of the class imbalance problem in convolutional neural networks},
  author={Buda, Mateusz and Maki, Atsuto and Mazurowski, Maciej A},
  journal={Neural networks},
  volume={106},
  pages={249--259},
  year={2018},
  publisher={Elsevier}
}

@article{snell2017prototypical,
  title={Prototypical networks for few-shot learning},
  author={Snell, Jake and Swersky, Kevin and Zemel, Richard},
  journal={Advances in neural information processing systems},
  volume={30},
  year={2017}
}

@article{field2007bootstrapping,
  title={Bootstrapping clustered data},
  author={Field, Christopher A and Welsh, Alan H},
  journal={Journal of the Royal Statistical Society Series B: Statistical Methodology},
  volume={69},
  number={3},
  pages={369--390},
  year={2007},
  publisher={Oxford University Press}
}

\section{Appendix}
\subsection{Baseline selection}
 We fixed the classifier before touching decoding. Nine architectures were trained on Kvasir-Capsule with weighted cross-entropy, official split (Tab.~\ref{tab:arch_baseline_kvasir}). DenseNet201+Swin-T won macro-F1 and MCC; DINOv2 ViT-L/14 led on rare-class recall and precision but trailed on the aggregate metrics, so DenseNet201+Swin-T went forward.

Table~\ref{tab:arch_baseline_kvasir} Architecture comparison, weighted CE, official split, single model, mean±sd over 3 seeds. No significance markers, these are independently trained models, not decoders on a shared frozen model.

\subsection{Loss and sampler selection}
With the architecture fixed, we swept loss and sampling strategy on that same backbone: weighted CE, focal loss, LDAM+DRW, and a class-balanced sampler, evaluated single-model and as a 3-seed ensemble (Tab.~\ref{tab:loss_sampler_kvasir}). Focal loss won macro-F1 and MCC in the single-model setting; weighted CE edges ahead on those same metrics in the ensemble setting. We selected focal loss anyway, as a follow-up check using decoder OT-alpha compared to raw argmax, confirmed focal to win on both F1 and MCC (Tab.~\ref{tab:loss_sampler_decoder_verify}).

\subsection{Class-balanced sampler ablation} 
We checked whether swapping the reported focal-loss/random-sampling recipe for a class-balanced sampler helps HCE or PACT further (Tab.~\ref{tab:ablation_cbsampler}). It does not, reliably: cbsampler's point-estimate rare recall looks higher in several rows, but that gain is driven by 2 of 6 seeds and is not significant in a direct per-video bootstrap comparison against focal (all $p \geq 0.44$), while macro-F1 drops significantly for HCE under cbsampler in the single-model setting. We report this as an honest negative result and keep focal as the main configuration.

\subsection{Conditioning granularity ablation} 
We tested whether conditioning on sub-video chunks (as small as $L=3$ frames) instead of the whole video changes anything, on both datasets (Tab.~\ref{tab:ablation_granularity}). It does not: macro-F1, MCC, rare recall, and rare precision all stay within noise of each other from whole-video conditioning down to $L=3$. We report this as a negative result too; whole-video conditioning is used everywhere else for simplicity, not because finer granularity was harmful.

\begin{figure*}[t]
  \centering
  \includegraphics[width=\textwidth]{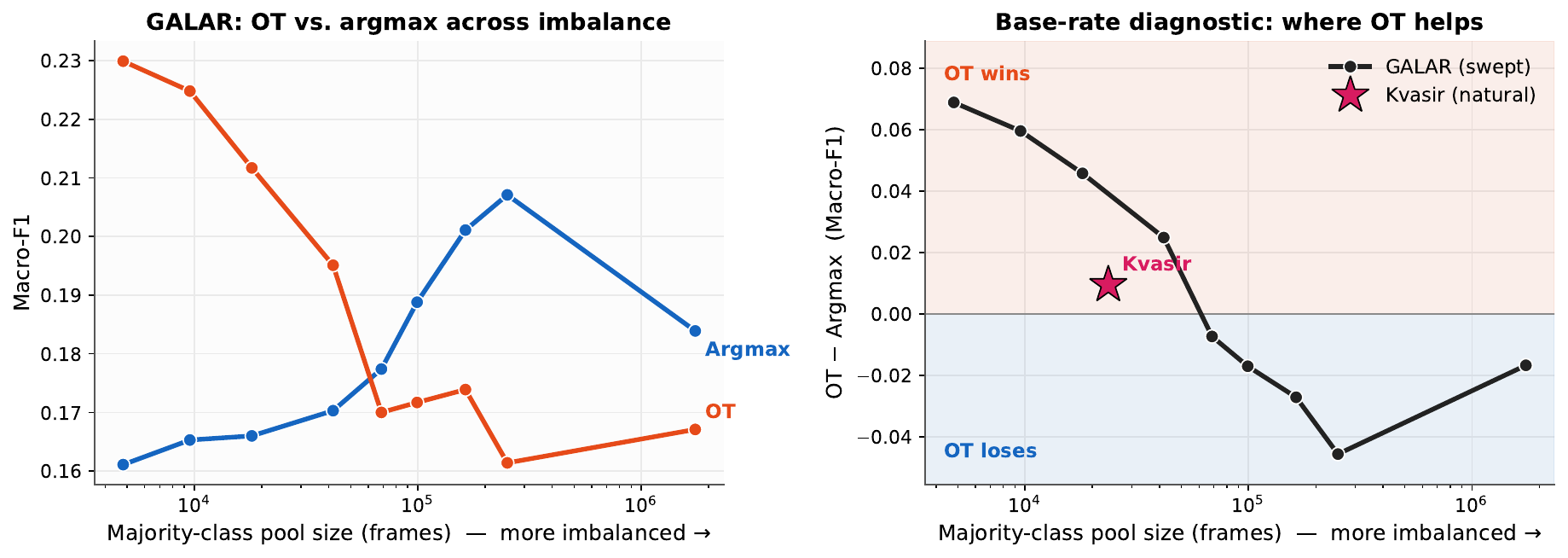}
  \caption{Base-rate diagnostic on GALAR: OT's macro-F1 gain over argmax as the majority-class pool size is swept (left: raw F1 curves; right: OT$-$argmax gain, with Kvasir-Capsule's natural, unswept imbalance level marked). The gain crosses from positive to negative as imbalance grows, from +0.07 F1 at small majority pools to $-0.046$ at large ones, including GALAR's own natural pool. Kvasir-Capsule's natural imbalance sits on the positive side of this crossover, explaining why OT helps on Kvasir but hurts on GALAR in the main results.}
  \label{fig:imbalance}
\end{figure*}

\paragraph{Imbalance diagnostic.} To characterize when OT actually helps, we swept GALAR's majority-class pool size and measured OT's macro-F1 gain over argmax at each level of imbalance (Tab.~\ref{fig:imbalance}). The gain crosses from positive to negative as the pool grows: OT beats argmax by up to +0.07 F1 at small majority pools, crosses zero between roughly 42k and 69k majority frames, and stays negative (down to $-0.046$) beyond that, including GALAR's own natural, uncapped pool. Kvasir-Capsule's natural imbalance level falls on the positive side of this crossover, which is exactly why OT holds up on Kvasir but hurts on GALAR in the main results.

\begin{figure*}[t]
  \centering
  \includegraphics[width=\textwidth]{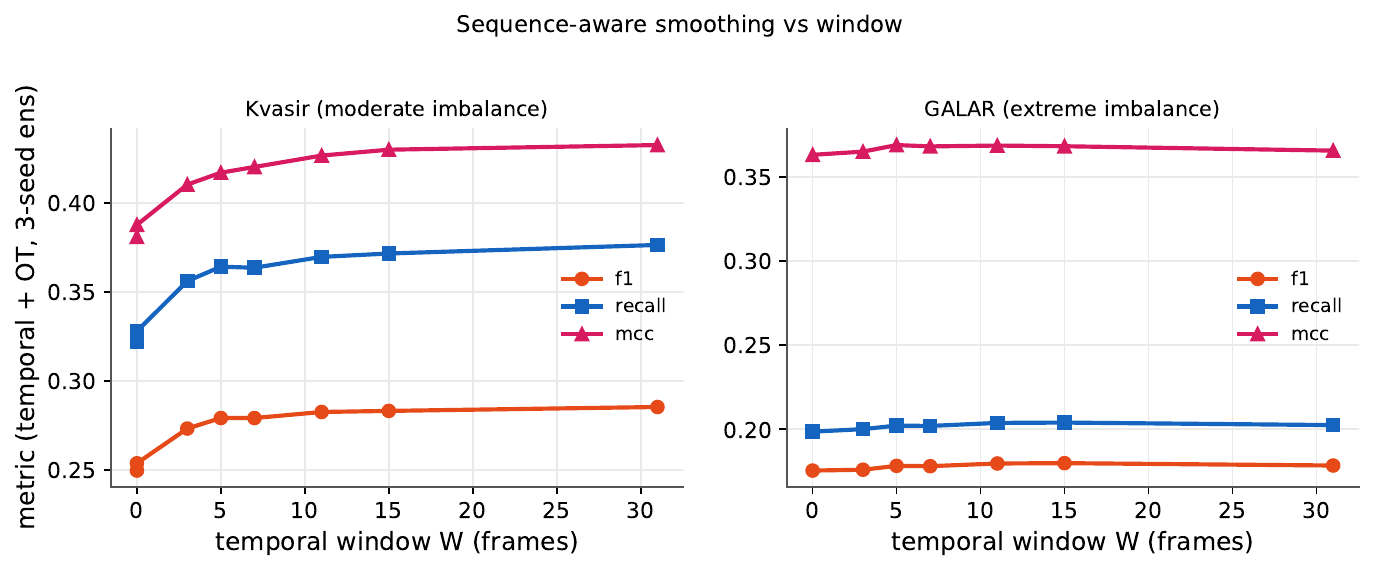}
  \caption{Temporal-smoothing window ablation under the temporal+OT ensemble decoder, Kvasir-Capsule (moderate imbalance) vs. GALAR (extreme imbalance). On Kvasir-Capsule, F1, recall, and MCC rise monotonically with window size $W$ (0.250 at $W{=}0$ to 0.286 at $W{=}31$). On GALAR the same sweep gains far less (0.176 to 0.180) and plateaus around $W{\approx}11$--$15$ before slightly declining by $W{=}31$, consistent with GALAR's extreme imbalance limiting how much sequence-level smoothing alone can recover.}
  \label{fig:temporal}
\end{figure*}

\paragraph{Temporal window ablation.} We swept the temporal-smoothing window $W$ under the temporal+OT ensemble decoder for both datasets (Tab.~\ref{fig:temporal}). On Kvasir-Capsule, F1, recall, and MCC all rise monotonically as $W$ grows, from 0.250 at $W{=}0$ to 0.286 at $W{=}31$, temporal context keeps helping the further out it looks. On GALAR, the same sweep gains far less (F1 0.176 to 0.180) and plateaus around $W{\approx}11$–$15$ before slightly declining by $W{=}31$, consistent with GALAR's extreme imbalance limiting how much sequence-level smoothing alone can recover.

\subsection{Experiment details}

\begin{enumerate}
    \item \textbf{Hardware and software.} All training ran on a single Tesla V100-SXM2-32GB GPU per job (node-locked via \texttt{nvidia.com/gpu.product} for run-to-run consistency), CUDA 12.1, cuDNN 8, PyTorch 2.2.0.
    \item \textbf{Optimizer and base recipe (all architectures).} AdamW, learning rate $10^{-4}$, weight decay 0.01, cosine learning-rate schedule, batch size 32, $224\times224$ input resolution.
    \item \textbf{Kvasir-Capsule training.} Up to 100 epochs, no early stopping; we keep the best-macro-F1 validation checkpoint rather than the final epoch's weights. Final recipe: focal loss ($\gamma=2.0$), random sampling, selected in Tables~\ref{tab:arch_baseline_kvasir}-~\ref{tab:loss_sampler_decoder_verify}.
    \item \textbf{GALAR training.} 30 epochs, early-stopping patience 8, batch size 64, mixed precision, majority classes capped at 40,000 training frames for tractability. Within each fold's video-disjoint test partition, frames are further uniformly subsampled to 10\% (label-blind, applied independently per split) purely to reduce evaluation compute; this does not change which videos are held out. Same focal loss / random sampling recipe as Kvasir.
    
    \item \textbf{Evaluation protocol.} 2 folds $\times$ 3 seeds per dataset, reported both as a single model and as a 3-seed ensemble. All decoder hyperparameters are selected by leave-video-out cross-validation only (no test-data access, fully inductive). Significance testing uses a per-video cluster bootstrap with a practical-significance floor (0.005 for F1/MCC, 0.003 for rare recall).
\end{enumerate}

\subsection{Theoretical background}

\subsection*{The macro-F1 plug-in and the base-rate wall}
Macro-F1 is non-decomposable, but at its optimum it behaves like a weighted linear metric with weights read off at that optimum (the linear-fractional/Frank-Wolfe argument of Parambath et al. 2014; Koyejo et al. 2014; Narasimhan et al. 2015). Writing $\text{macro-F1} = \frac{1}{K}\sum_c 2\,TP_c/(P_c+T_c)$ and holding $(P_c, T_c)$ at their optimal values, the maximizer is a weighted argmax, $\arg\max_c a_c\,p_c(x)$ with $a_c = 2/(P_c+T_c)$, i.e., an additive per-class offset $\delta_c = \log a_c$. Since $P_c = TP_c/\text{prec}_c$, $T_c = TP_c/\text{rec}_c$, and $TP_c = \text{rec}_c\, T_c = \text{rec}_c\, N\pi_c$, substituting gives
$$\delta_c = -\log \pi_c + \underbrace{\log\!\left(\frac{2\,\text{prec}^*_c}{\text{prec}^*_c+\text{rec}^*_c}\right)}_{=:g_c} + \text{const}.$$
Two consequences follow. First, logit adjustment (Menon et al. 2021, $\delta_c=-\tau\log\pi_c$) is the special case $g_c=\text{const}$; under real imbalance $g_c$ is not constant, near $\log 2$ for well-separated head classes and strongly negative once a class's precision has collapsed, which is why logit adjustment over-boosts the rare tail and costs MCC. Second, the base-rate wall falls out of the same optimum: for a rare class at fixed discriminability, $\text{prec}(r) = r\pi/(r\pi+\text{FPR}(r)(1-\pi)) \to r\pi/\text{FPR}(r)$ as $\pi\to 0$, so precision becomes linear in $\pi_c$ in the tail and $g_c \to \log\pi_c + \text{const}$. The correction then exactly cancels the $-\log\pi_c$ boost: below the wall, the macro-F1-optimal rule stops boosting the rare class at all, with no threshold or tuned constant needed to produce that behavior, it is forced by the optimum itself.

\subsection*{Why EGA has the right form}
The plug-in rule above is a fixed point, not a closed-form formula ($\delta_c$ depends on $P_c$ at the optimum), and naive iteration on it diverges. EGA (Expected Gain Arbitration) is a stable solver for this fixed point via greedy coordinate ascent on the exact expected macro-F1 effect of reassigning one frame. Since $F1_c = 2\,TP_c/(P_c+T_c)$,
$$\frac{\partial F1_c}{\partial TP_c} = \frac{2(T_c + FP_c)}{(P_c + T_c)^2}, \qquad \frac{\partial F1_c}{\partial FP_c} = -\frac{2\,TP_c}{(P_c+T_c)^2},$$
so with a calibrated posterior $p_c(x)$, moving frame $x$ from its current class $b$ to class $c$ has expected effect
$$\mathbb{E}[\Delta](x, b{\to}c) = \big[p_c\,g^+_c + (1-p_c)\,g^-_c\big] - \big[p_b\,g^+_b + (1-p_b)\,g^-_b\big],$$
and the rule moves $x$ only when this is positive, taking the best $c$. Counts $(TP_c, P_c, T_c)$ are estimated on validation under the base decoder and iterated twice so counts reflect earlier reassignments. This also identifies logit adjustment precisely as EGA's $g_c=\text{const}$ special case, matching the plug-in derivation above.

A closely related recent approach, Schultheis et al. (NeurIPS 2023), optimizes an expected-test-utility objective for long-tail extreme multi-label classification via block-coordinate ascent, the same general strategy EGA uses. EGA differs in operating per-frame within a video-conditioned decoding pipeline rather than over a fixed label set, and in solving a simpler two-pass greedy fixed point rather than their general regret-bounded framework; we adopt the simpler solver because our setting composes with the video-level conditioning offset in Section~3.4, which their framework does not address.

\subsection*{Escaping the wall: video-level conditioning}
Every offset-based decoder above treats class prevalence $\pi_c$ as one number, which is what puts rare-disease frame classification below the wall. For VCE specifically, frame-level prevalence factorizes:
$$\pi_{\text{frame}}(c) = \underbrace{P(\text{patient has } c)}_{\sim 0.02} \times \underbrace{P(\text{frame shows } c \mid \text{patient has } c)}_{\sim 0.005\text{--}0.05} \approx 10^{-4}.$$
The disease is rare across patients, not within one: a patient with a finding has many frames showing it. The two factors are each decided far above the wall (at $\sim\!200\times$ and $\sim\!500\times$ the frame rate respectively), and aggregating frames within a video suppresses noise, lowering the false-positive rate at which each factor is estimated. The product of the two prevalences is conserved, but the difficulty is not, since precision depends on both prevalence and FPR, and aggregation improves both.

Critically, this leaves the family the plug-in theorem bounds: the theorem constrains decoders of the form $\arg\max_c[\log p_c + \delta_c]$ with a single global $\delta_c$. Conditioning $\delta_c$ on the patient's own video $v$,
$$\arg\max_c\big[\log p_c(x) + \delta_c(v)\big],$$
falls outside that family. Concretely, with $r_c = P(\text{frame shows } c \mid \text{video contains } c)$ estimated from validation videos containing $c$, and $q_c(v)$ a presence model applied to a top-$k$ aggregate of $p_c$ over $v$'s frames,
$$\pi_c(v) = q_c(v)\,r_c + (1-q_c(v))\,\alpha\pi_c, \qquad \delta_c(v) = \lambda \log\!\big(\pi_c(v)/\pi_c\big),$$
with $\alpha, \lambda$ selected on validation macro-F1. As $q_c(v)\to 1$ this gives a large positive boost $\log(r_c/\pi_c)$; as $q_c(v)\to 0$ it suppresses $c$ in that video. Because the offset only fires where video-level evidence supports the disease, and actively suppresses it elsewhere, precision is protected in a way uniform rare-class boosting cannot achieve, uniform boosting raises recall by firing in every video and destroys precision as a result.

\subsection*{Optimal transport (Sinkhorn) baseline}
The OT baseline solves an entropic optimal-transport assignment via Sinkhorn's algorithm: given per-frame log-posteriors and a target class marginal (the empirical class prior raised to a tuned power $\alpha$), it iterates dual potentials $u, v$ in log-space,
\begin{equation}
\begin{split}
u &\leftarrow \log r - \text{logsumexp}_c(G + v), \\
v &\leftarrow \log c_{\text{marg}} - \text{logsumexp}_n(G + u),
\end{split}
\end{equation}

where $G$ is the temperature-scaled log-posterior matrix and $r$ is the uniform row marginal, then assigns each frame to its row-argmax under the resulting transport plan. This re-matches the predicted class proportions to a chosen target marginal, correcting global frequency bias without touching the underlying posteriors' rank order within a frame. An unbalanced variant relaxes the column (class) constraint by a softness $\kappa\in[0,1]$, recovering hard OT at $\kappa=1$ and plain argmax at $\kappa=0$ as special cases, so a confidently-normal frame is not forced to satisfy a rare-class quota it has no evidence for.

\subsection*{Temporal smoothing}
Temporal smoothing averages each frame's posterior over a within-video neighborhood of $\pm W$ frames (never crossing a video boundary, using only model posteriors and frame indices, never labels). The rationale is structural: true lesions span contiguous runs of frames, while artifact false positives tend to be isolated spikes. Averaging the sequence suppresses the spikes while preserving the runs, raising precision without discarding any frame.

\subsection{Rescue confusion matrices}
\begin{figure*}[t]
  \centering
  \includegraphics[width=\linewidth]{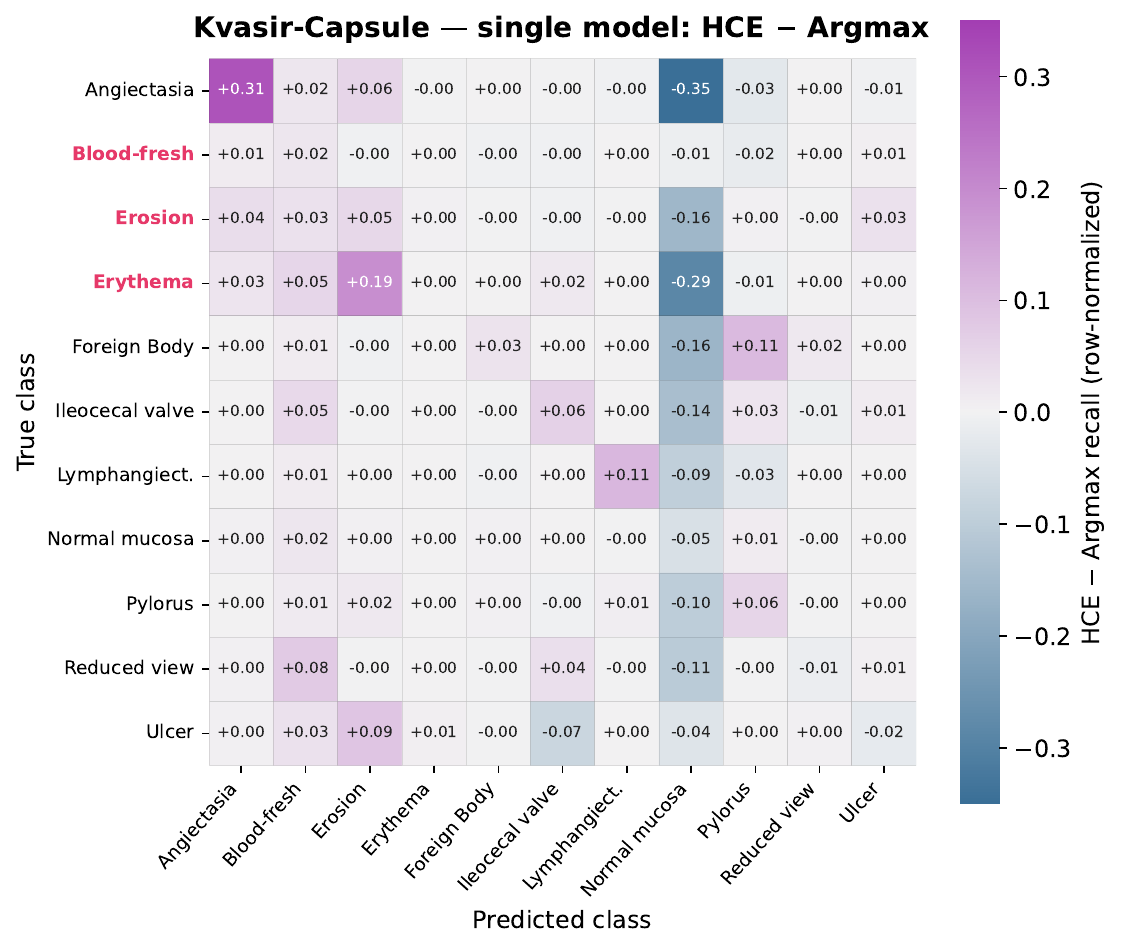}
  \caption{HCE minus argmax, row-normalized confusion-matrix delta, Kvasir-Capsule single model. Diagonal = per-class recall change. Angiectasia (+0.308), Lymphangiectasia (+0.115), Ileocecal valve (+0.064), and Pylorus (+0.056) gain the most; Erosion (rare) gains +0.052. Blood-fresh and Erythema (rare) stay at zero recall under both methods, Blood-fresh has posterior AUC 0.22, below chance, unrecoverable by any decoder. Normal mucosa's own recall drops slightly ($-0.048$).}
  \label{fig:confmat_rescue_kvasir_single}
\end{figure*}

\begin{figure*}[t]
  \centering
  \includegraphics[width=\linewidth]{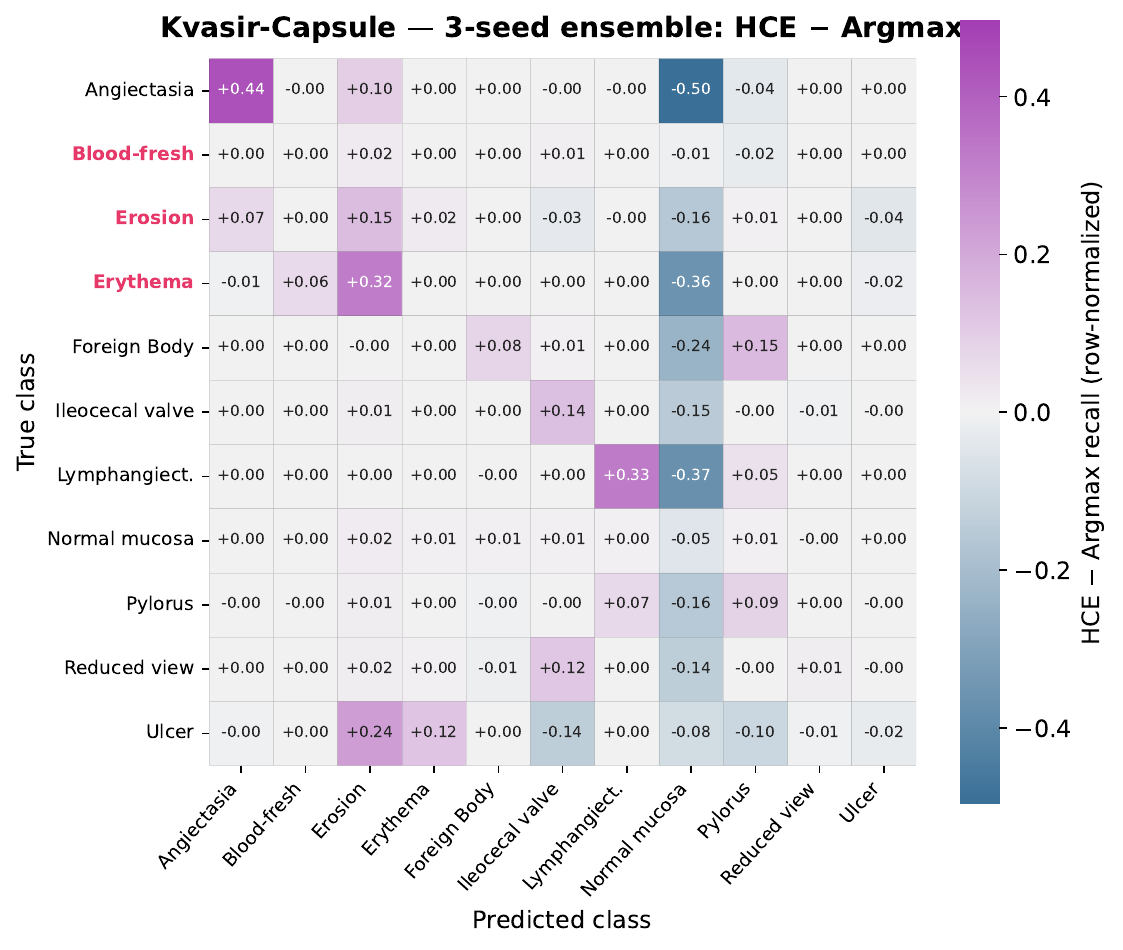}
  \caption{HCE minus argmax, row-normalized confusion-matrix delta, Kvasir-Capsule 3-seed ensemble. Same pattern as the single-model matrix, amplified: Angiectasia (+0.443), Lymphangiectasia (+0.326), Ileocecal valve (+0.137), Pylorus (+0.086), Erosion (rare, +0.147). Blood-fresh and Erythema remain at zero recall. Normal mucosa drops $-0.049$.}
  \label{fig:confmat_rescue_kvasir_ens}
\end{figure*}

\begin{figure*}[t]
  \centering
  \includegraphics[width=\linewidth]{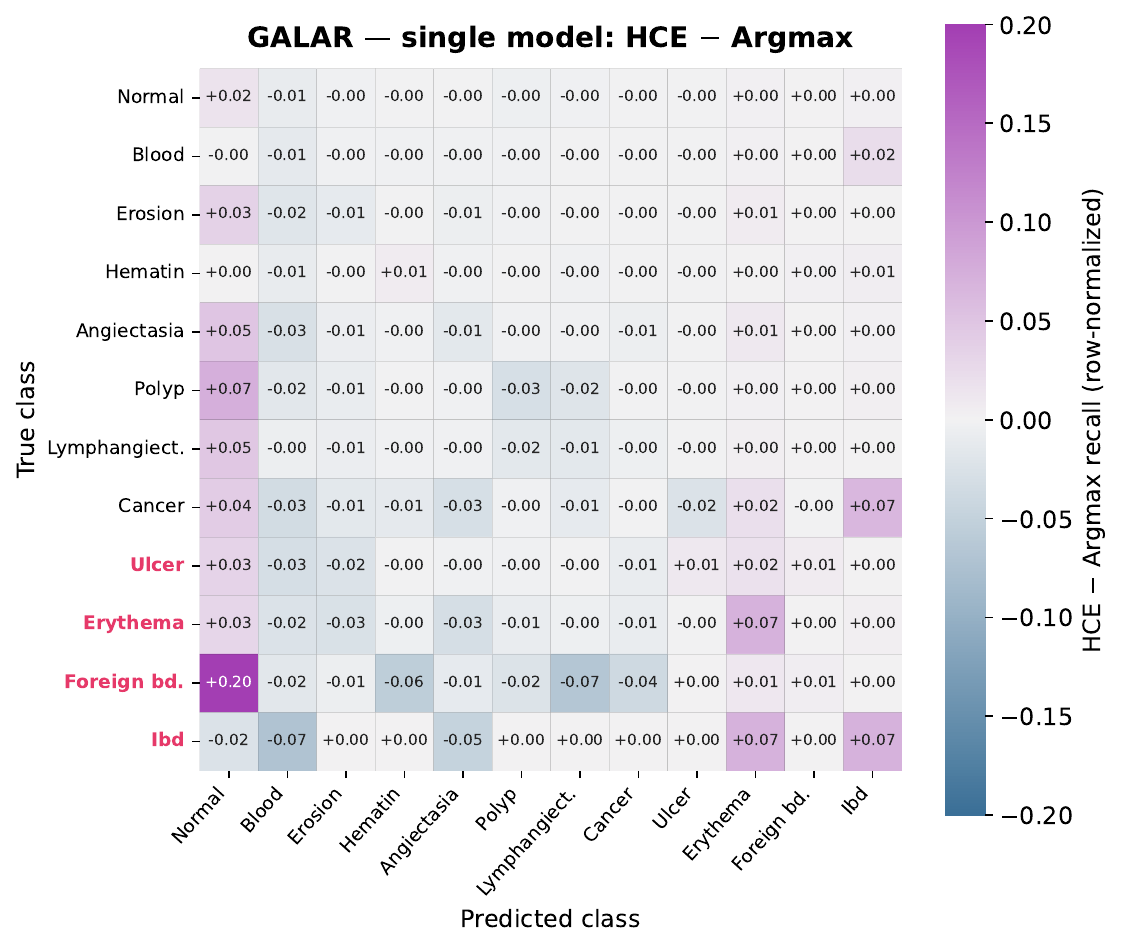}
  \caption{HCE minus argmax, row-normalized confusion-matrix delta, GALAR single model. Erythema (+0.071) and ulcer (+0.011) gain recall; IBD moves from 0 to 0.071, but on only 42 test frames, a single-frame effect rather than a reliable signal. Polyp ($-0.030$), angiectasia ($-0.015$), lymphangioectasis ($-0.015$), erosion ($-0.013$), and blood ($-0.014$) lose recall. Normal's own recall rises slightly (+0.016).}
  \label{fig:confmat_rescue_galar_single}
\end{figure*}

Across all three matrices (Figs.~\ref{fig:confmat_rescue_kvasir_single} to~\ref{fig:confmat_rescue_galar_single}), the largest gains come from classes argmax was defaulting to the majority class on: on Kvasir-Capsule (single model), Angiectasia's and Lymphangiectasia's argmax predictions were mostly ``Normal mucosa'' (56.2\% and 58.3\% of their frames), and HCE recovers most of that directly (down to 21.2\% and 48.8\%), so the same mechanism, pulling frames back out of ``Normal'', accounts for almost the entire single-model gain there (Angiectasia recall $+0.308$, Lymphangiectasia $+0.115$). GALAR runs the opposite direction: its cost classes' lost recall moves mostly \emph{into} normal rather than out of it (e.g., polyp $-0.030$, angiectasia $-0.015$), while normal's own recall rises only slightly ($+0.016$). GALAR's imbalance is far more extreme, so the base-rate wall bites harder there, and HCE's video-level conditioning fixes fewer of the head-class-vs-normal confusions that it resolves so cleanly on Kvasir-Capsule.

\subsection*{Complete Bootstrap Intervals}
Table~\ref{tab:ci_full} reports the full set of per-video cluster bootstrap intervals behind every significance statement in the paper: five metrics, four evaluation settings, and both baselines, for forty cells in total. Each entry pairs the point-estimate difference taken from the results tables with the 95\% interval obtained by resampling whole test videos with replacement. We separate the two deliberately. The bootstrap's own resampled mean can drift from the point estimate when the two folds disagree in sign, most visibly for rare recall against optimal transport on Kvasir ensembles, where the resampled mean is $+0.048$ against a point estimate of $+0.019$; the interval is what the bootstrap is for, and the point estimate is what the results tables report. Five cells resolve, all on GALAR. Against temporal smoothing the resolved gains are in rare recall ($+0.037$ $[+0.010, +0.068]$ single-model, $+0.040$ $[+0.005, +0.067]$ ensemble); against optimal transport they are in macro F1 ($+0.022$ and $+0.025$) and in MCC on the single-model setting ($+0.094$ $[+0.004, +0.168]$), while no rare-recall comparison against optimal transport resolves. Two further cells fall just outside: rare recall against temporal smoothing on Kvasir ensembles has a lower bound at zero, and MCC against optimal transport on GALAR ensembles at $-0.001$. Kvasir intervals are uniformly the wider ones, which the design explains: each fold holds 25 test videos and a given rare class is present in roughly seven of them, so resampling at the video level has little to work with. The GALAR recall gains are not offset by confirmed costs elsewhere, as rare precision against temporal smoothing is negative in point estimate ($-0.022$ and $-0.042$) but neither interval excludes zero.

\begin{table*}[t]
  \centering
  \begin{tabular}{lccccc}
    \toprule
    Architecture & Macro-F1 & Macro Recall & MCC & Rare Recall & Rare Precision \\
    \midrule
    DenseNet-161 & 0.2291$\pm$0.0144 & 0.2710$\pm$0.0127 & 0.3450$\pm$0.0198 & 0.0385$\pm$0.0120 & 0.0149$\pm$0.0042 \\
    DenseNet201 + Swin-T & \textbf{0.2383$\pm$0.0065} & \textbf{0.2848$\pm$0.0096} & \textbf{0.3619$\pm$0.0075} & 0.0539$\pm$0.0098 & 0.0223$\pm$0.0042 \\
    DINOv2 ViT-L/14 & 0.1992$\pm$0.0114 & 0.2361$\pm$0.0201 & 0.2855$\pm$0.0063 & \textbf{0.0763$\pm$0.0400} & \textbf{0.0903$\pm$0.0885} \\
    DINOv2 ViT-S/14 & 0.2010$\pm$0.0057 & 0.2405$\pm$0.0129 & 0.2857$\pm$0.0131 & 0.0689$\pm$0.0480 & 0.0214$\pm$0.0069 \\
    EfficientNet-B2 & 0.2251$\pm$0.0213 & 0.2697$\pm$0.0201 & 0.3446$\pm$0.0355 & 0.0366$\pm$0.0201 & 0.0165$\pm$0.0054 \\
    EfficientNet-B4 & 0.2231$\pm$0.0104 & 0.2547$\pm$0.0125 & 0.3384$\pm$0.0155 & 0.0308$\pm$0.0133 & 0.0244$\pm$0.0159 \\
    FocalConvNet & 0.2116$\pm$0.0066 & 0.2391$\pm$0.0037 & 0.3411$\pm$0.0142 & 0.0452$\pm$0.0161 & 0.0148$\pm$0.0055 \\
    ResNet-152 & 0.2142$\pm$0.0042 & 0.2665$\pm$0.0047 & 0.3441$\pm$0.0047 & 0.0534$\pm$0.0223 & 0.0715$\pm$0.0981 \\
    ViT-Small/16 & 0.2193$\pm$0.0113 & 0.2459$\pm$0.0095 & 0.3549$\pm$0.0174 & 0.0216$\pm$0.0069 & 0.0203$\pm$0.0208 \\
    \bottomrule
  \end{tabular}
  \caption{Kvasir-Capsule architecture comparison, weighted CE, official split, single model. Mean$\pm$sd over 3 seeds (2 folds averaged per seed). Rare = Blood - fresh, Erosion, Erythema. \textbf{Bold} = best per column. No significance testing (compares independently trained models, not decoders on a shared frozen model; see main results tables for per-video cluster bootstrap significance). DenseNet201+Swin-T is carried forward to the loss/sampling sweep.}
  \label{tab:arch_baseline_kvasir}
\end{table*}

\begin{table*}[t]
  \centering
  \begin{tabular}{lccccc}
    \toprule
    Loss / Sampler & Macro-F1 & Macro Recall & MCC & Rare Recall & Rare Precision \\
    \midrule
    \multicolumn{6}{l}{\textit{Single model}} \\
    \midrule
    Weighted CE & 0.2383$\pm$0.0065 & \textbf{0.2848$\pm$0.0096} & 0.3619$\pm$0.0075 & 0.0539$\pm$0.0098 & 0.0223$\pm$0.0042 \\
    Focal loss & \textbf{0.2403$\pm$0.0076} & 0.2664$\pm$0.0088 & \textbf{0.3683$\pm$0.0195} & 0.0511$\pm$0.0126 & 0.0307$\pm$0.0199 \\
    LDAM+DRW & 0.2370$\pm$0.0161 & 0.2806$\pm$0.0160 & 0.3636$\pm$0.0247 & 0.0337$\pm$0.0070 & 0.0187$\pm$0.0005 \\
    Class-balanced sampler + CE & 0.2290$\pm$0.0125 & 0.2798$\pm$0.0212 & 0.3678$\pm$0.0189 & \textbf{0.0959$\pm$0.0417} & \textbf{0.0589$\pm$0.0462} \\
    \midrule
    \multicolumn{6}{l}{\textit{3-seed ensemble}} \\
    \midrule
    Weighted CE & \textbf{0.2513$\pm$0.0004} & \textbf{0.2910$\pm$0.0309} & \textbf{0.3829$\pm$0.0231} & 0.0543$\pm$0.0618 & 0.0239$\pm$0.0208 \\
    Focal loss & 0.2379$\pm$0.0327 & 0.2604$\pm$0.0074 & 0.3753$\pm$0.0253 & 0.0399$\pm$0.0360 & 0.0186$\pm$0.0153 \\
    LDAM+DRW & 0.2388$\pm$0.0372 & 0.2804$\pm$0.0021 & 0.3759$\pm$0.0620 & 0.0316$\pm$0.0077 & 0.0192$\pm$0.0220 \\
    Class-balanced sampler + CE & 0.2296$\pm$0.0086 & 0.2743$\pm$0.0314 & 0.3690$\pm$0.0409 & \textbf{0.0917$\pm$0.1147} & \textbf{0.0376$\pm$0.0156} \\
    \bottomrule
  \end{tabular}
  \caption{Kvasir-Capsule loss/sampling comparison, DenseNet201+Swin-T, official split. Single model: mean$\pm$sd over 3 seeds (2 folds averaged per seed). Ensemble: mean$\pm$sd over 2 folds (n=2, seed variation already collapsed by ensembling). \textbf{Bold} = best per column in each block. No significance testing (see main results tables for per-video cluster bootstrap significance). Rare = Blood - fresh, Erosion, Erythema. LA-loss and CRT were also tried but have no ensemble run and are omitted. Focal loss is carried forward to decoding.}
  \label{tab:loss_sampler_kvasir}
\end{table*}

\begin{table*}[t]
  \centering
  \begin{tabular}{lcccc}
    \toprule
    Loss / Sampler & Argmax & LA & OT-alpha & OT-reg \\
    \midrule
    Weighted CE & 0.2383/0.3619 & 0.2384/0.3620 & 0.2477/0.3730 & 0.2445/0.3681 \\
    Focal loss & \textbf{0.2403}/\textbf{0.3683} & \textbf{0.2403}/\textbf{0.3683} & \textbf{0.2499}/\textbf{0.3908} & \textbf{0.2485}/\textbf{0.3899} \\
    LDAM+DRW & 0.2370/0.3636 & 0.2370/0.3636 & 0.2463/0.3804 & 0.2460/0.3804 \\
    Class-balanced sampler + CE & 0.2290/0.3678 & 0.2290/0.3678 & 0.2448/0.3797 & 0.2435/0.3783 \\
    \bottomrule
  \end{tabular}
  \caption{Kvasir-Capsule, DenseNet201+Swin-T: Macro-F1/MCC under each decoder, single model, official split (2 folds averaged, no seed-level sd shown). \textbf{Bold} = best config per decoder per metric (F1 or MCC independently). Verifies the focal-loss choice against the decoder actually used for selection (OT-alpha), not just raw argmax: focal wins both F1 and MCC under OT-alpha even though it is not the argmax-best config.}
  \label{tab:loss_sampler_decoder_verify}
\end{table*}

\begin{table*}[t]
  \centering
  \begin{tabular}{lccccc}
    \toprule
    Config & Macro-F1 & Macro Recall & MCC & Rare Recall & Rare Precision \\
    \midrule
    \multicolumn{6}{l}{\textit{Single model}} \\
    \midrule
    HCE, focal & \textbf{0.2654$\pm$0.0413} & 0.2986$\pm$0.0311 & \textbf{0.3955$\pm$0.0573} & 0.0635$\pm$0.0345 & 0.0191$\pm$0.0130 \\
    HCE, cbsampler & 0.2509$\pm$0.0306$^{\downarrow}$ & \textbf{0.3093$\pm$0.0553} & 0.3804$\pm$0.0355 & \textbf{0.1534$\pm$0.1724} & \textbf{0.0446$\pm$0.0437} \\
    PACT, focal & \textbf{0.2571$\pm$0.0360} & 0.2806$\pm$0.0207 & \textbf{0.3924$\pm$0.0537} & 0.0484$\pm$0.0306 & 0.0178$\pm$0.0130 \\
    PACT, cbsampler & 0.2370$\pm$0.0225$^{\downarrow}$ & \textbf{0.2870$\pm$0.0422} & 0.3839$\pm$0.0332 & \textbf{0.0899$\pm$0.1234} & \textbf{0.0769$\pm$0.1186} \\
    \multicolumn{6}{l}{\textit{3-seed ensemble}} \\
    \midrule
    HCE, focal & \textbf{0.3023$\pm$0.0243} & 0.3472$\pm$0.0047 & \textbf{0.4421$\pm$0.0569} & 0.1013$\pm$0.0578 & 0.0224$\pm$0.0112 \\
    HCE, cbsampler & 0.2792$\pm$0.0125 & \textbf{0.3760$\pm$0.0566} & 0.4195$\pm$0.0702 & \textbf{0.2493$\pm$0.2300} & \textbf{0.0269$\pm$0.0062} \\
    PACT, focal & 0.2536$\pm$0.0188 & 0.2907$\pm$0.0157 & \textbf{0.4003$\pm$0.0227} & 0.0645$\pm$0.0510 & 0.0167$\pm$0.0073 \\
    PACT, cbsampler & \textbf{0.2567$\pm$0.0026} & \textbf{0.3338$\pm$0.0550} & 0.3994$\pm$0.0463 & \textbf{0.2097$\pm$0.2020} & \textbf{0.0329$\pm$0.0164} \\
    \bottomrule
  \end{tabular}
  \caption{Ablation: class-balanced sampler (cbsampler) vs. focal on Kvasir, HCE/PACT. \textbf{Bold} = numerically higher of the pair per column (not a significance claim). $\uparrow$/$\downarrow$ marks a DIRECT per-video bootstrap comparison between the two configs (p$<$0.05 and $|\Delta|\geq$0.003-0.005), not either config's own significance vs. its temporal baseline. cbsampler's higher point-estimate rare recall is driven by 2 of 6 outlier seeds and is not significant in this direct comparison (all p$\geq$0.44); reported as an honest negative/inconclusive ablation, not a win. Focal remains the reported main-result configuration.}
  \label{tab:ablation_cbsampler}
\end{table*}

\begin{table*}[t]
  \centering
  \begin{tabular}{lcccc}
    \toprule
    Granularity & Macro-F1 & MCC & Rare Recall & Rare Precision \\
    \midrule
    \multicolumn{5}{l}{\textit{Kvasir-Capsule, single model}} \\
    \midrule
    Whole video (L=$\infty$) & 0.2535$\pm$0.0336 & 0.3903$\pm$0.0387 & 0.0467$\pm$0.0339 & 0.0190$\pm$0.0116 \\
    L=1000 & 0.2538$\pm$0.0338 & 0.3906$\pm$0.0383 & 0.0470$\pm$0.0341 & 0.0193$\pm$0.0119 \\
    L=300 & 0.2556$\pm$0.0361 & 0.3912$\pm$0.0381 & 0.0459$\pm$0.0352 & 0.0173$\pm$0.0106 \\
    L=100 & 0.2531$\pm$0.0352 & 0.3905$\pm$0.0451 & 0.0437$\pm$0.0322 & 0.0173$\pm$0.0109 \\
    L=30 & 0.2558$\pm$0.0377 & 0.4081$\pm$0.0459 & 0.0431$\pm$0.0297 & 0.0155$\pm$0.0119 \\
    L=10 & 0.2563$\pm$0.0341 & 0.3961$\pm$0.0515 & 0.0408$\pm$0.0305 & 0.0159$\pm$0.0090 \\
    L=3 & 0.2500$\pm$0.0324 & 0.3977$\pm$0.0483 & 0.0418$\pm$0.0280 & 0.0182$\pm$0.0117 \\
    \multicolumn{5}{l}{\textit{GALAR, single model}} \\
    \midrule
    Whole video (L=$\infty$) & 0.1862$\pm$0.0151 & 0.4424$\pm$0.0846 & 0.0248$\pm$0.0167 & 0.0673$\pm$0.0814 \\
    L=1000 & 0.1871$\pm$0.0166 & 0.4387$\pm$0.0847 & 0.0283$\pm$0.0198 & 0.0610$\pm$0.0705 \\
    L=300 & 0.1866$\pm$0.0168 & 0.4354$\pm$0.0822 & 0.0280$\pm$0.0192 & 0.0610$\pm$0.0711 \\
    L=100 & 0.1859$\pm$0.0170 & 0.4319$\pm$0.0815 & 0.0300$\pm$0.0202 & 0.0577$\pm$0.0662 \\
    L=30 & 0.1863$\pm$0.0175 & 0.4294$\pm$0.0814 & 0.0314$\pm$0.0215 & 0.0558$\pm$0.0624 \\
    L=10 & 0.1863$\pm$0.0171 & 0.4268$\pm$0.0831 & 0.0366$\pm$0.0319 & 0.0478$\pm$0.0508 \\
    L=3 & 0.1860$\pm$0.0165 & 0.4331$\pm$0.0843 & 0.0254$\pm$0.0244 & 0.0529$\pm$0.0573 \\
    \bottomrule
  \end{tabular}
  \caption{Ablation: video-level vs. sub-video-chunk conditioning granularity (L = frames per conditioning unit). All metrics stay within noise of each other across $L=\infty$ (whole video) down to $L=3$ on both datasets: granularity is inert on real data, the video-level unit is not doing meaningful work relative to any finer split. Reported as a negative result; whole-video conditioning is used throughout the rest of the paper for simplicity, not because finer granularity was harmful.}
  \label{tab:ablation_granularity}
\end{table*}

\begin{table*}[t]
  \centering
  \footnotesize
  \setlength{\tabcolsep}{4pt}
  \begin{tabular}{lcccc}
    \toprule
    Metric & Kv-S & Kv-E & GA-S & GA-E \\
    \midrule
    \multicolumn{5}{l}{\textit{HCE $-$ temporal smoothing}} \\
    \quad Rare recall & +0.026 [-0.008, +0.066] & +0.072 [+0.000, +0.179] & \textbf{+0.037 [+0.010, +0.068]} & \textbf{+0.040 [+0.005, +0.067]} \\
    \quad Rare precision & -0.002 [-0.033, +0.026] & +0.005 [-0.023, +0.033] & -0.022 [-0.047, +0.007] & -0.042 [-0.095, +0.008] \\
    \quad Rare F1 & +0.008 [-0.009, +0.030] & +0.015 [-0.002, +0.052] & +0.009 [-0.009, +0.018] & +0.011 [-0.028, +0.032] \\
    \quad Macro F1 & +0.022 [-0.024, +0.038] & +0.067 [-0.012, +0.088] & +0.002 [-0.005, +0.005] & +0.002 [-0.011, +0.009] \\
    \quad MCC & +0.014 [-0.036, +0.061] & +0.063 [-0.012, +0.134] & -0.007 [-0.017, +0.003] & -0.003 [-0.008, +0.002] \\
    \midrule
    \multicolumn{5}{l}{\textit{HCE $-$ optimal transport}} \\
    \quad Rare recall & +0.016 [-0.033, +0.114] & +0.019 [-0.077, +0.195] & +0.029 [-0.009, +0.071] & +0.016 [-0.011, +0.045] \\
    \quad Rare precision & -0.022 [-0.106, +0.085] & -0.003 [-0.098, +0.053] & +0.024 [-0.019, +0.057] & +0.039 [-0.037, +0.088] \\
    \quad Rare F1 & +0.007 [-0.015, +0.041] & +0.017 [-0.010, +0.066] & +0.008 [-0.006, +0.022] & +0.004 [-0.021, +0.029] \\
    \quad Macro F1 & +0.017 [-0.054, +0.060] & +0.040 [-0.048, +0.081] & \textbf{+0.022 [+0.006, +0.036]} & \textbf{+0.025 [+0.003, +0.044]} \\
    \quad MCC & +0.006 [-0.076, +0.089] & +0.034 [-0.052, +0.117] & \textbf{+0.094 [+0.004, +0.168]} & +0.114 [-0.001, +0.213] \\
    \bottomrule
  \end{tabular}
  \caption{Per-video cluster bootstrap for HCE against both baselines, all metrics and all four settings (Kv/GA = Kvasir-Capsule/GALAR, S/E = single/3-seed ensemble). Each cell gives the point-estimate difference followed by the 95\% interval. The point estimate is the difference of the means reported in the results tables; the bootstrap resamples whole videos and supplies the interval only, since the two diverge when folds disagree in sign. \textbf{Bold} marks intervals excluding zero, which is the significance criterion used throughout. Against temporal smoothing the resolved gains are in rare recall on both GALAR settings; against optimal transport they are in macro F1 on both GALAR settings and MCC on GALAR single-model, while no rare-recall comparison against optimal transport resolves at this sample size.}
  \label{tab:ci_full}
\end{table*}

\end{document}